\documentclass[preprint]{aastex}
\usepackage{psfig}
\begin{document}
\title{X-ray Energy Spectra of the Super-soft X-ray Sources CAL87 and RXJ0925.7--4758 Observed with ASCA}
\author{Ken Ebisawa\altaffilmark{1}, Koji Mukai\altaffilmark{1}, Taro Kotani\altaffilmark{2}
}
\affil{Laboratory for High Energy Astrophysics, NASA/GSFC, Greenbelt, MD, 20771, USA}
\altaffiltext{1}{code 662,  also Universities Space Research Association}
\altaffiltext{2}{code 661,  also National Research Council}
\author{Kazumi Asai, Tadayasu Dotani, Fumiaki Nagase
}
\affil{Institute of Space and Astronautical Science, Yoshinodai, 
Sagamihara, Kanagawa, 229-8510 Japan}
\author{H. W. Hartmann, J. Heise}
\affil{
SRON Laboratory for Space Research,  Sorbonnelaan 2, NL-3584 CA Utrecht, The Netherlands}
\author{P. Kahabka\altaffilmark{3}}
\affil{Astronomical Institute and Center for High Energy Astrophysics, University of Amsterdam, Kruislaan 403, 1098 SJ Amsterdam, The Netherlands}
\altaffiltext{3}{Present address: 
Sternwarte, Universitaet Bonn, Auf dem Huegel 71, 53121 Bonn, Germany
}
\and
\author{A. van Teeseling}
\affil{Universit\"ats-Sternwarte, Geismarlandstr. 11, 37083,
G\"otteingen, Germany}

\begin{abstract}
We report observation results of the super-soft X-ray sources (SSS) 
CAL87 and RXJ0925.7--4758 with the
X-ray CCD cameras (Solid-state Imaging  Spectrometer; SIS) on-board the 
ASCA satellite.
Because of the superior energy resolution of SIS ($\Delta E/E \sim 10 \% $
at 1 keV) relative to previous instruments,  we could study
detailed X-ray spectral structure of these  sources for the first time.
We have applied theoretical spectral models to CAL87, and
constrained the white dwarf mass and intrinsic luminosity as 
$0.8 - 1.2 M_\odot $ and $4 \times 10^{37}$--$1.2 \times 10^{38}$
erg s$^{-1}$, respectively.  However, we have found the observed luminosity
is an order of magnitude smaller than the theoretical estimate, 
which indicates the white
dwarf is permanently blocked by the accretion disk, and
we are observing a scattering emission by a fully ionized 
accretion disk corona (ADC) whose column density is $\sim 1.5 \times 10^{23}$ cm$^{-2}$.
Through simulation, we have shown that the orbital eclipse can be explained 
by the ADC model, such that a part of the extended X-ray emission from the ADC
is blocked by the companion star filling its Roche lobe.

We have found that  very high surface gravity and temperature, $\sim 10^{10}$
cm s$^{-2}$ and $\sim $100 eV respectively, 
as well as a strong absorption edge at $\sim$ 1.02 keV, 
are required
to explain the X-ray energy spectrum of RXJ0925.7--4758.
These values are  only possible for an extremely heavy white dwarf near the Chandrasekhar
limit. 
 Although  the super soft source luminosity should 
 be $\sim 10^{38}$ erg s$^{-1}$ at the Chandrasekhar limit, 
the observed luminosity of  RXJ0925.7--4758 is nearly two orders of
magnitude smaller even assuming  an extreme  distance of $\sim$
10 kpc. To explain the luminosity discrepancy,  
we propose a model that very thick 
matter which was previously ejected from the system,  as a form of jets, 
intervenes  the line of sight, and reduces the luminosity significantly
due to Thomson scattering. 

\end{abstract}


\section{Introduction}

Dozens of the Super Soft X-ray Sources (SSS), which  characteristically are
 radiating most energies in the softest X-ray energy band ($\lesssim$ 0.5 keV),
have been discovered in LMC, SMC, M31 and Milky Way by  {\em Einstein} and ROSAT 
(for review, e.g., Kahabka and van den Heuvel 1997).
It is believed that most SSS are white dwarf binaries with large mass accretion rates
($\sim 1 -5 \times 10^{-7} M_\odot$ yr$^{-1}$), and the energy source is
steady nuclear burning 
on the white dwarf surface (van den Heuvel et al.\ 1992; Heise et al.\ 1994).
Observationally, our knowledge on the X-ray energy spectra of SSS has been limited by the capability of 
the conventional X-ray detectors.  While the energy bands of the {\em Einstein} IPC and ROSAT PSPC (0.1 -- 2 keV) 
are suitable for the study of SSS, 
their energy resolutions ($\Delta E/E \gtrsim 1$ at 0.5 keV) have not allowed one to clearly identify 
any possible spectral structures such as emission lines or absorption edges.
Having a superior energy resolution ($\Delta E/E \sim 10 \%$ at 0.5 keV), 
the ASCA Solid State Spectrometer
(SIS; Tanaka, Inoue and Holt 1994) is potentially a powerful instrument for spectroscopic study of SSS, though its detectable 
energy band (0.4  -- 10 keV) is slightly too high to 
cover the typical SSS spectra.

The two SSS, CAL87 and RXJ0925.7--475, are known to have relatively
hard energy spectra, thereby we have chosen these two sources for 
precise spectral study with the ASCA SIS.  
While most SSS have characteristic blackbody temperatures of $\lesssim$ 30 eV, 
those of CAL87 and RXJ0925.7--475 are $\gtrsim$ 40 eV (Schmidtke et al.\ 1993; 
Motch, Hasinger and Pietsch 1994).  
CAL87 was discovered during the LMC survey observations with the {\em Einstein} IPC (Long, Helfand 
and Grabelsky 1981).
It is an eclipsing binary with a  10.6 hour orbital period, exhibiting orbital eclipses in the optical band and 
accompanying shallow X-ray dips (Schmidtke et al.\ 1993; Alcock et al.\ 1997; Asai et al.\ 1998). 
With the assumption that the non-degenerate companion is a F-star, 
Cowley et al.\ (1990) speculated that the compact object is massive and may be a black hole.
However, HST observations revealed two adjacent stars at 0.\arcsec88 and 0.\arcsec65 from CAL87
(Deutsch et al.\ 1996), which have yet to be taken into account to determine the 
optical properties  of CAL87.
Schandl, Meyer-Hofmeister and Meyer (1997) successfully  modeled the observed optical light curve
 of CAL87
assuming a 0.75 $M_\odot$ white-dwarf and a 1.5  $M_\odot$ companion.
To account for the shallow X-ray dips, the Accretion Disk Corona (ADC) model was proposed in which 
the blocking material partially covers the largely extended X-ray emitting corona (Schmidtke et al.\ 1993).

RXJ0925.7--475 was discovered in the ROSAT Galactic Plane survey project  (Motch, Hasinger and Pietsch 1994). 
This  unusual, heavily absorbed star exhibits
most X-ray emission at energies above $\sim$ 0.5 keV,
while most other SSS have little X-rays above $\sim$ 0.5 keV (Motch, Hasinger and Pietsch 1994; Motch 1996). 
The orbital period is $\sim$ 3.8 days, and the optical counterpart indicates  strong reddening (Motch, Hasinger and Pietsch 1994; Motch 1996;
Schmidtke et al.\ 2000). A transient optical jet, similar to 
those observed from other super-soft sources, has been also observed
 (Motch 1998).   
The reddening, strength of the interstellar absorption lines, and ROSAT spectra are all 
consistent with a large hydrogen column density of $N_H \gtrsim 10^{22}$ cm$^{-2}$, 
which suggests that the source is located
behind the nearby ($d$=425 pc) Vela Sheet molecular cloud (Motch, Hasinger and Pietsch 1994).
Optical light curve analysis suggests that mass of the compact object
is in the range of 0.5 -- 1.7 $M_\odot$ and that of the donor 
star is in 1 -- 2 $M_\odot$
(Schmidtke et al.\ 2000).

In this paper, we report precise analysis of the  
X-ray energy spectra of CAL87 and RXJ0925.7--4758 observed with ASCA.
Owing to the superior energy resolution of the ASCA SIS, we could reveal 
detailed X-ray spectral features of the two sources.  
The ASCA CAL87 light curve exhibits 
shallow orbital eclipses, and its energy spectrum show
deep absorption edges which are identified as \ion{O}{7} and \ion{O}{8}
edges, as already reported by  Asai et al.\ (1998). Preliminary spectral
analysis of RXJ0925.7--4758 has been reported by Ebisawa et al.\ (1996), and
similar ASCA spectral analysis has been also carried out by Shimura (2000).
As an extension of these studies,
we apply theoretical  LTE (Local Thermal Equilibrium) and Non-LTE white-dwarf 
atmospheric spectral models to CAL87 and RXJ0925.7--4758, 
and constrain the white dwarf parameters.  Also, we carry out a simulation to
explain the CAL87 orbital light curve in the framework of the  accretion disk corona 
(ADC) model,
such that a part of the extended ADC X-ray emission  is blocked by a  companion star
and observed as a shallow and wide  orbital eclipse.



\section{Observations}

RXJ0925.7--475 was observed with ASCA from 13:28UT on December 22 1994 to 02:02UT on December 23.
CAL87 was observed from 06:08UT on September 6, 1996 to 08:30UT on September 8.
ASCA carries two SIS (SIS0 and SIS1) and two GIS (Gas Imaging Spectrometers;
GIS2 and GIS3) detectors, each combined with an X-ray telescope,
and the four sensors  pointing the same direction (Tanaka, Inoue 
and Holt 1994). 
Since RXJ0925.7--475 and CAL87 do not emit significantly in the GIS energy band (0.7--10 keV),
we could not detect enough photons with GIS to carry out a spectral analysis.
Therefore, we present only the SIS spectral analysis results in this paper. The GIS data of
RXJ0925.7--475 was used only for a coherent pulsation search, but
we obtained null results in the frequency range $2 \times 10^{-4}$ to 8  Hz.
After appropriate SIS data 
screening, the total exposure time was 15 ksec for RXJ0925.7--475 and 70 ksec for CAL87.
The SIS 1CCD FAINT mode was used for both observations, and we 
applied the standard instrumental calibration.
On each detector, the  X-ray source photons were accumulated within $\sim$ 4 arcmin from 
the center of the source.
Having made sure that both SIS detectors gave identical results, we combined the data and responses 
from the two sensors  for further analysis to achieve  better statistics.
The background spectra were made from the region on the same chips where source photons are negligible,
normalized by the selected area, and subtracted from the source spectra.  Spectral fitting
analysis was carried out with XSPEC spectral fitting package (Arnaud 1996).
For CAL87, we used essentially the same dataset as in Asai et al.\ (1998).
Average count rates of the sources 
after background subtraction are 0.015 cts/s/SIS for CAL87 (orbit averaged) 
and 0.28 cts/s/SIS for RXJ0925.7--475, respectively,  in 0.4 -- 2.0 keV.

\section{Data Analysis}

\subsection{CAL87}

As reported by Asai et al.\ (1998), ASCA data indicates shallow ($\sim 60 \%$)
and wide ($\sim 30 \%$ of the period) X-ray dips  in phase
with the optical eclipse of the white dwarf, which  confirms the 
ROSAT results (Schmidtke et al.\ 1993; Kahabka, Pietsch and Hasinger 1994).
In order to study spectral variation,
we extracted spectra from the eclipse and out of the eclipse period
in addition to the entire averaged spectrum.
We took the ephemeris and orbital period from Alcock et al.\ (1997);
the X-ray eclipse center (orbital phase $\phi$=0) is MJD=50331.9103 and the period is 0.44267714 day.
For the eclipse period, we took  the orbital period $-0.16 < \phi < 0.16$
(see the orbital light curve, Fig.\ 2 in Asai et al.\ 1998), outside of which 
is considered the non-eclipse period.

We  take into account 
both the interstellar absorption in the Milkyway 
and an intrinsic absorption within the LMC.
In the spectral model fitting, 
following Asai et al.\ (1998), we fix the former column density 
at $8 \times 10^{20}$ cm$^{-2}$ and allow the latter one to be free.
We assume the cosmic abundance for the Milkyway absorption,
and half the metal abundance for the LMC absorption
(Vassiliadis and Wood 1994; Dennefeld 1989).
In the Tables below, we show only the amount of intrinsic absorption in the LMC 
thus determined.

\subsubsection{Spectral Analysis with Blackbody + Edges}\label{cal87_simple}
First, to characterize the observed spectrum, we analyze the energy spectrum 
applying a  simple phenomenological model,
and consider how much we can tell  about the source
with minimum assumptions.
We use the average spectrum taken from the entire period.

As we have shown in Asai et al. (1998), the CAL87 spectral fit
requires a deep absorption edge at $\sim$ 0.85 keV, which may be
interpreted due to  \ion{O}{7} and \ion{O}{8} edges
at 0.74 keV and 0.87 keV respectively; these two edges are so close
and the source flux is low, ASCA SIS cannot resolve them.
We fit the spectrum with these absorption edges with the fixed energies.
The optical depth of the \ion{O}{8} edge is not constrained, so we fixed it
to 10.0, which practically allow no photons to escape above the edge
energy.
In this model, normalization of the continuum (= effective emission
area), temperature, and  amount of absorption are
strongly correlated, and  hardly  constrained  (Table 1, left column).

\begin{figure}
\centerline{
\psfig{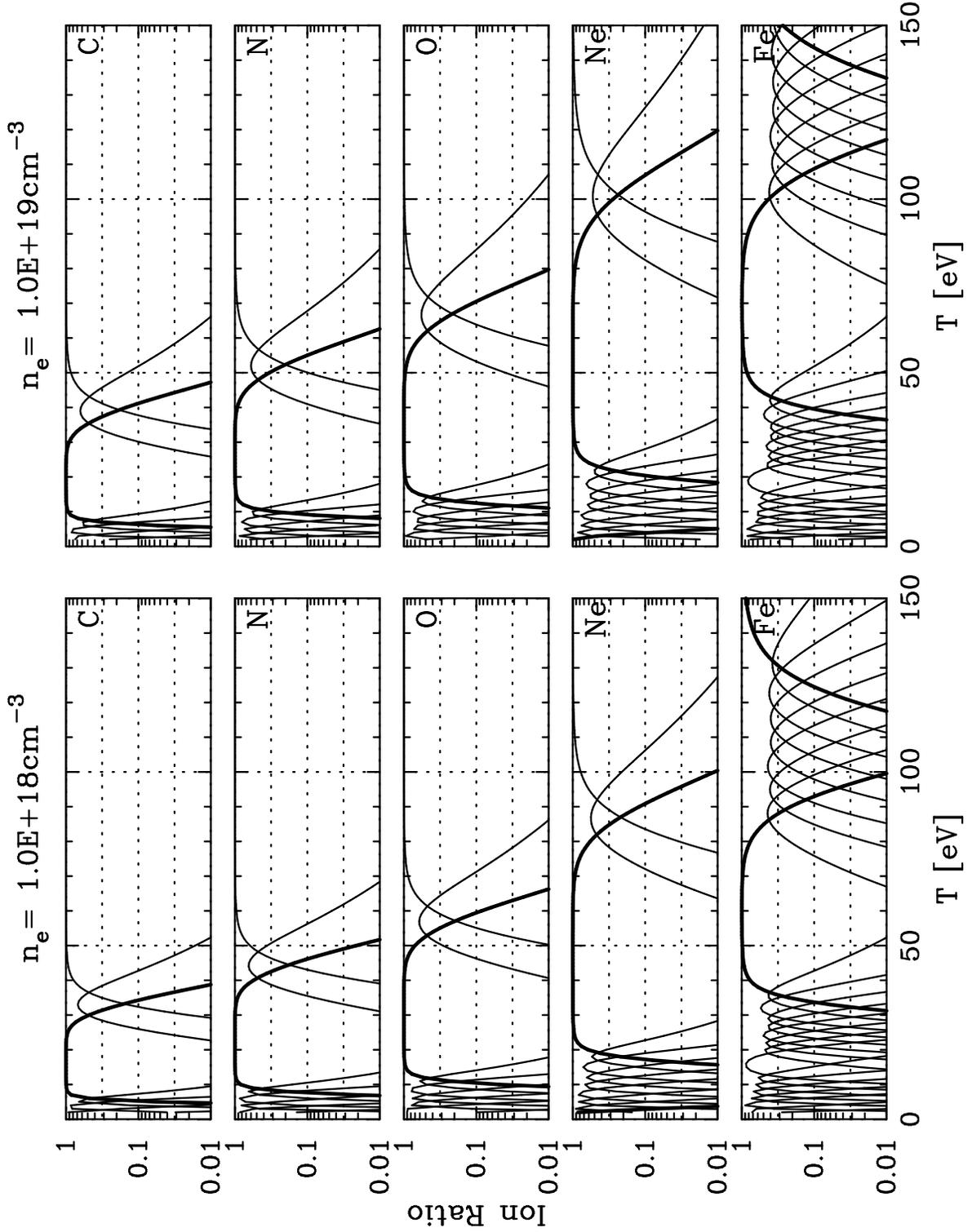}
}
\caption{
Ion Fractions of the major heavy elements as a function of the temperature
for the two electron densities 
$10^{18}$ and $10^{19}$ cm$^{-3}$.  LTE condition is assumed.
Fractions of the helium-like ions and neon-like ion (for iron)
 are indicated with the thick lines.
}\label{ion_fraction}
\end{figure}

We may constrain the temperature based on a simple physical consideration,
from the fact that the \ion{O}{7} and \ion{O}{8} edges are observed simultaneously.
If we assume the LTE condition, the ionization balance is determined by the temperature, $T$,  and the electron density, 
$N_{\rm e}$, through the Saha relation.
The electron density can be  related with the temperature and the pressure by the 
equation of state 
$n_e \approx P/\: k\: T$.  
Neglecting the radiation pressure, the gas 
 pressure may be estimated from the hydrostatic equilibrium condition 
as $P \approx g/\kappa_{\rm R}$, where $g$ is the surface gravity, which 
is typically $10^8$ to $10^9$ cm s$^{-2}$ for white dwarfs,
and $\kappa_{\rm R}$ is the Rosseland mean opacity, $\sim$ 0.4 cm$^{2}$ g$^{-1}$.  
From these relations, for a typical temperature of several tens of eV, 
the 
electron densities  in the white dwarf atmosphere of an optical depth of $\sim$
unity are considered to be  in the range of $10^{18}$  to $10^{19}$ cm$^{-3}$.
In Figure \ref{ion_fraction}, we show the 
ion ratios for C, N, O,  Ne and Fe  calculated from the Saha relation
as functions of the temperature,  for the two electron densities $n_e=  10^{18}$  and $10^{19}$ cm$^{-3}$.
It is seen that the condition that the \ion{O}{7} and \ion{O}{8} edges are both
present (with \ion{O}{8} being  more abundant)
 requires    $T\approx$ 60 -- 80 eV.

\begin{figure}
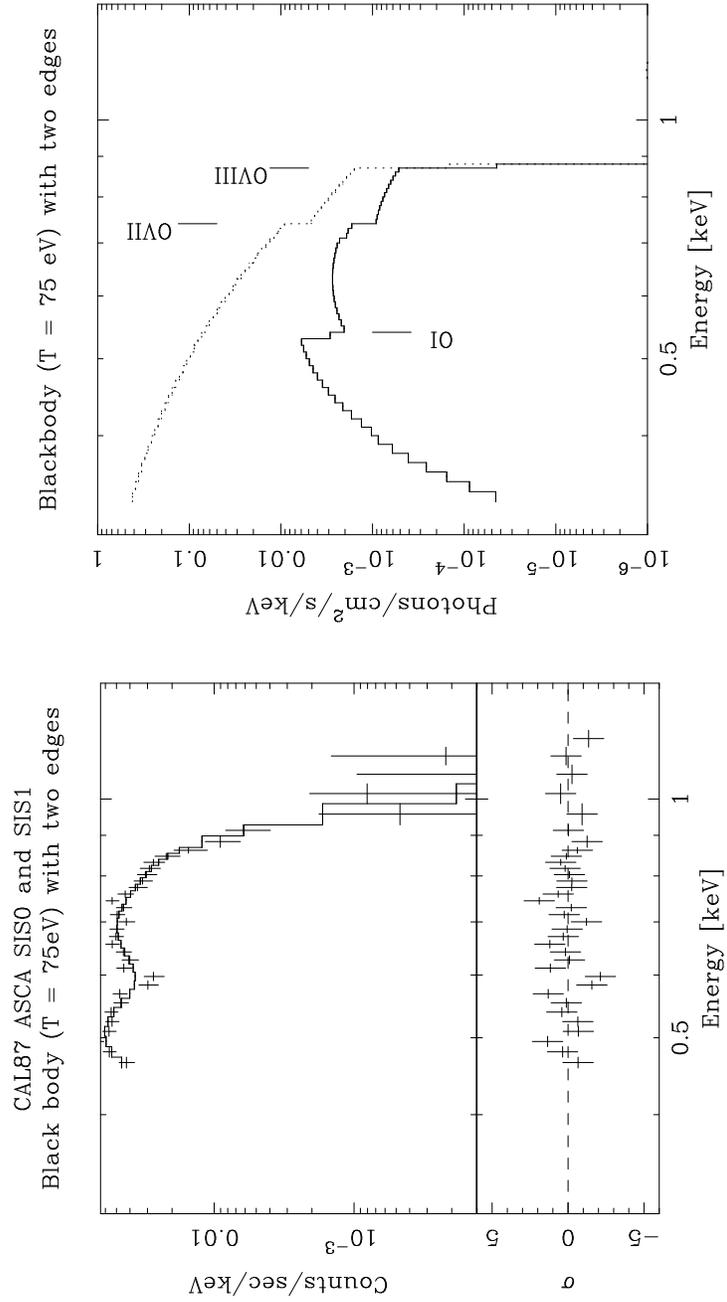

\centerline{
\psfig{figure=f2b.eps,%
      height=9.cm,bbllx=0.cm,bblly=4.cm,bburx=21.cm,bbury=21cm,%
      angle=0,clip=}}
\centerline{
\psfig{figure=f2a.eps,%
      height=9.cm,bbllx=0.cm,bblly=4.cm,bburx=21.cm,bbury=21cm,%
      angle=0,clip=}
}
\caption{Blackbody ($T$=75 eV) plus two edge  model fit for CAL87. 
Left: Observed spectrum and best-fit model convolved with the detector response.
Residuals from the best-fit model are shown in the lower panel.
Right: The unfolded best-fit model spectra
are shown with (solid line) and without (dashed line)
taking into account the interstellar absorption.
}\label{bb_edge}
\end{figure}

Therefore we fix the blackbody  temperature at 75 eV. 
The results are shown in Figure \ref{bb_edge} and Table \ref{bb_edge_table} 
(right column).
We found the radius and luminosity with $T=75 $ eV are  reasonable
for a massive ($\gtrsim$ 1 $M_\odot$) white dwarf (section \ref{white_dwarf}; 
Figure \ref{white_dwarf_figure}).
Thus,  based on a simple model, we see 
that the energy spectrum of CAL87 is likely to be
explained by atmospheric emission from a massive white dwarf.
Our data suggest that the energy spectrum emitted from 
the white-dwarf atmosphere,  where the strong absorption
edges are produced,  is directly seen.
Consequently,  if the white dwarf is permanently blocked from the line-of-sight
by outer part of the accretion disk (Schandl, Meyer-Hofmeister and Meyer 1997) and/or 
the X-rays are coming from extended ADC
(Schmidtke et al.\ 1993), the corona has to be optically
thin so that the scattering does not smear the absorption edges.

\subsubsection{Theoretical White-dwarf Models}\label{cal87_white_dwarf_model}
As shown in the previous section, it is likely that the energy
spectrum of CAL87 is explained by emission from a hot white-dwarf atmosphere.
Therefore, we next apply theoretical white-dwarf spectral models 
to constrain  the white-dwarf parameters.

We adopt  the same LTE (local thermal equilibrium) model used by
Heise, van Teeseling and Kahabka (1994) and 
van Teeseling, Heise and Kahabka (1996) to fit the SSS energy spectra,  
and the NLTE (non-LTE) model by  Hartmann and Heise (1997).
In both models, 
plane parallel geometry and hydrostatic equilibrium are assumed, 
and  energy spectra have been
calculated for  given surface gravities, $g$,  and effective temperatures, $T_{eff}$.  In the NLTE model, 
the photoionization effect is additionally taken into account 
to calculate the ionization balance.  In the LTE model,
all the ions of H, He, C, N, O, Ne, Mg, Si,
P, Ar, Ca, Fe and Ni are taken into account, and all the edges
are included.  In the NLTE model,
only major ions and atomic levels of H, He, C, N, O and Ne are included
(see Hartmann and Heise 1997 for detail).

Complex line opacities and line blanketing are not taken 
into account in both models.
Significance of  the line opacities as well as the non-LTE effects 
in the emerging spectra has  been studied by several authors such as
Rauch (1997), Hartmann et al.\ (1999) and Barman et al.\ (2000).
The NLTE model not including the line opacities (the same model used
in this paper) results in the absorption edges at  higher energies
compared to the LTE model with the same effective temperature, as well
as  a significant amount of the high-energy flux
 (Hartmann and
Heise 1997).  Additionally including the line opacity to the
NLTE model reduces the edge energies and high-energy flux
 (Hartmann et al.\ 1999),
but not down to the same level as the LTE model (Barman et al.\ 2000).
Therefore, although there are systematic differences among the NLTE models by 
different authors, the ``correct'' model spectrum is expected
to lie between the LTE and NLTE model spectra
we use to fit the observed spectra.
In that sense, 
we will be presenting conservative limits of the white dwarf parameters
in the present paper.

To fit the observed data with these models
for a given surface gravity, the best-fit temperature is
searched for by interpolating one dimensional temperature grids
on which model spectra have been  precalculated;
the temperature grids are every 5 $\times 10^4$ K 
from 5 $\times 10^5$ K to $10^6$ K 
for the LTE model, and 
every  $10^4$ K from 3 $\times 10^5$ K to $10^6$ K for the NLTE model.
The models mildly depend on the surface gravity, but
the temperature and gravity are hardly determined 
simultaneously (see section \ref{cal87_gravity_abundance}).
We adopt a surface gravity of $10^9$ cm s$^{-2}$,
which is appropriate for a massive white dwarf with a mass $\gtrsim 1 M_\odot$;
validity of this assumption is  checked a posteriori (section \ref{white_dwarf}).
The models also weakly depend on the elemental abundances.
For CAL87, the LMC abundance in which metalicity 
is a quarter to half of the cosmic abundance
is used,  either taken from 
Vassiliadis and Wood (1994) (LTE model) or Dennefeld (1989) (NLTE model).
Effects of changing the surface gravity and abundances are discussed in
section \ref{cal87_gravity_abundance}  and \ref{rxj_gravity_abundance}.

\begin{figure}
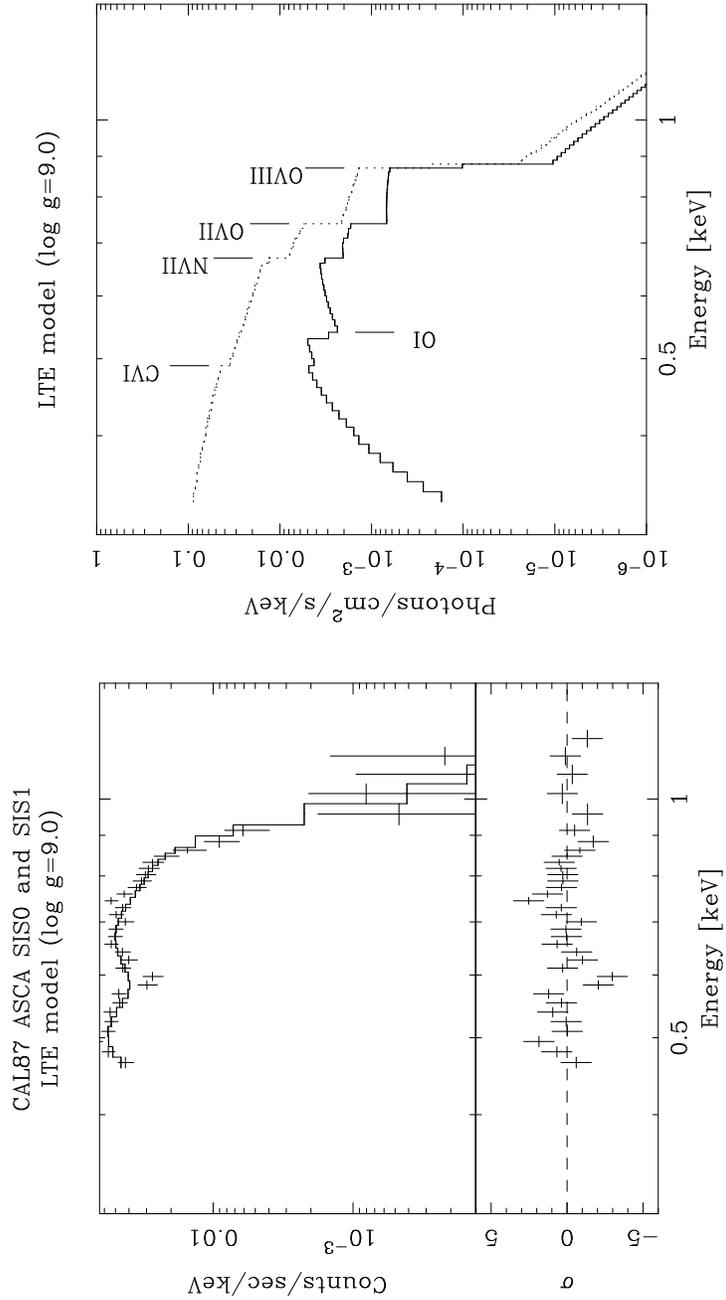

\centerline{
\psfig{figure=f3b.eps,%
      height=9.cm,bbllx=0.cm,bblly=4.cm,bburx=21.cm,bbury=21cm,%
      angle=0,clip=}}
\centerline{
\psfig{figure=f3a.eps,%
      height=9.cm,bbllx=0.cm,bblly=4.cm,bburx=21.cm,bbury=21cm,%
      angle=0,clip=}
}
\caption{LTE model fit for CAL87.  Left: Observed spectrum and the best-fit model 
convolved with the detector response.  
Right: The best-fit model before and after the interstellar absorption.
Absorption edges in the model and those due to the interstellar absorption are indicated.  
Note that energies of the \protect\ion{O}{8} and \ion{Ne}{1} edges coincide (0.87 keV). 
}\label{cal87_lte_fit}
\end{figure}

In Table \ref{cal87_lte_fit_table} and Figure \ref{cal87_lte_fit}, we show the
results of the LTE model 
fit with $g=10^9$ cm s$^{-2}$.
The best-fit temperature, normalization (= effective emission area) and hydrogen column density  have been
determined for the average, eclipse, and out of the eclipse spectra.
To fit the eclipse and out of the eclipse spectra,  the  temperature and
either column density or normalization are fixed to the best-fit
values determined from the average spectrum, since these parameters are not 
constrained simultaneously. 
Confidence contour map of the hydrogen column density and 
the effective emission area is shown in Figure 
\ref{cal87_contour}.  

The model gives acceptable fits.
The temperature is tightly constrained, as expected from 
the condition that both \ion{O}{7} and \ion{O}{8} are present  (section \ref{cal87_simple}).  
In addition to the \ion{O}{7} and \ion{O}{8} edges, the best-fit model shows
\ion{N}{7} and \ion{C}{6} edges (see Figure \ref{cal87_lte_fit}).  
Although the 
temperature is  constrained, the normalization and the hydrogen column density are 
strongly correlated (Figure \ref{cal87_contour}), and the 
effective emission area is uncertain by about a factor of two.  
This is because the energy band of ASCA SIS is 
above $\sim$ 0.5 keV, where model spectra are not
very sensitive to slight change of the column densities.  
Contemporary instruments having both better energy resolution
and  sensitivity below $\sim$ 0.5 keV,
such as the grating spectrometers on Chandra or XMM-Newton,
will be able to constrain the column density and normalization
simultaneously.
In section \ref{white_dwarf}, we will 
compare the observed luminosity and radius with those 
predicted by theoretical calculations, and discuss the white dwarf parameters.
The spectral change during the eclipse is explained either 
due to an increase of the hydrogen column density by $\sim 5 \times 10^{20} $cm$^{-2}$ or a
decrease of the
emission area by $\sim$ 20 \% (see section \ref{cal87_eclipse};  also see 
Asai et al.\ 1998).

\begin{figure}
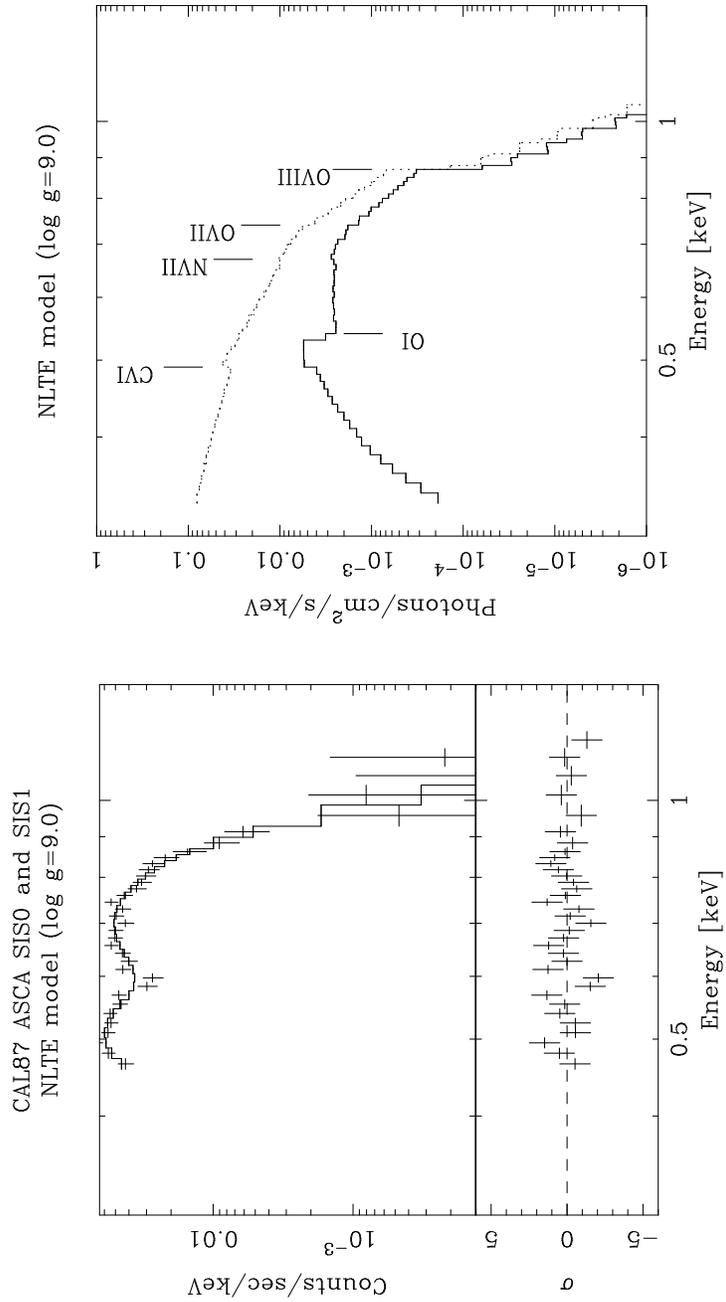

\centerline{
\psfig{figure=f4b.eps,%
      height=9.cm,bbllx=0.cm,bblly=4.cm,bburx=21.cm,bbury=21cm,%
      angle=0,clip=}}
\centerline{
\psfig{figure=f4a.eps,%
      height=9.cm,bbllx=0.cm,bblly=4.cm,bburx=21.cm,bbury=21cm,%
      angle=0,clip=}
}
\caption{NLTE model fit for CAL87.  Left: Observed spectrum and the best-fit model 
convolved with the detector response.  
Right: The best-fit model before and after the interstellar absorption.
Absorption edges in the model and those due to the interstellar absorption are indicated.  
}\label{cal87_nlte_fit}
\end{figure}

\begin{figure}
\centerline{
\psfig{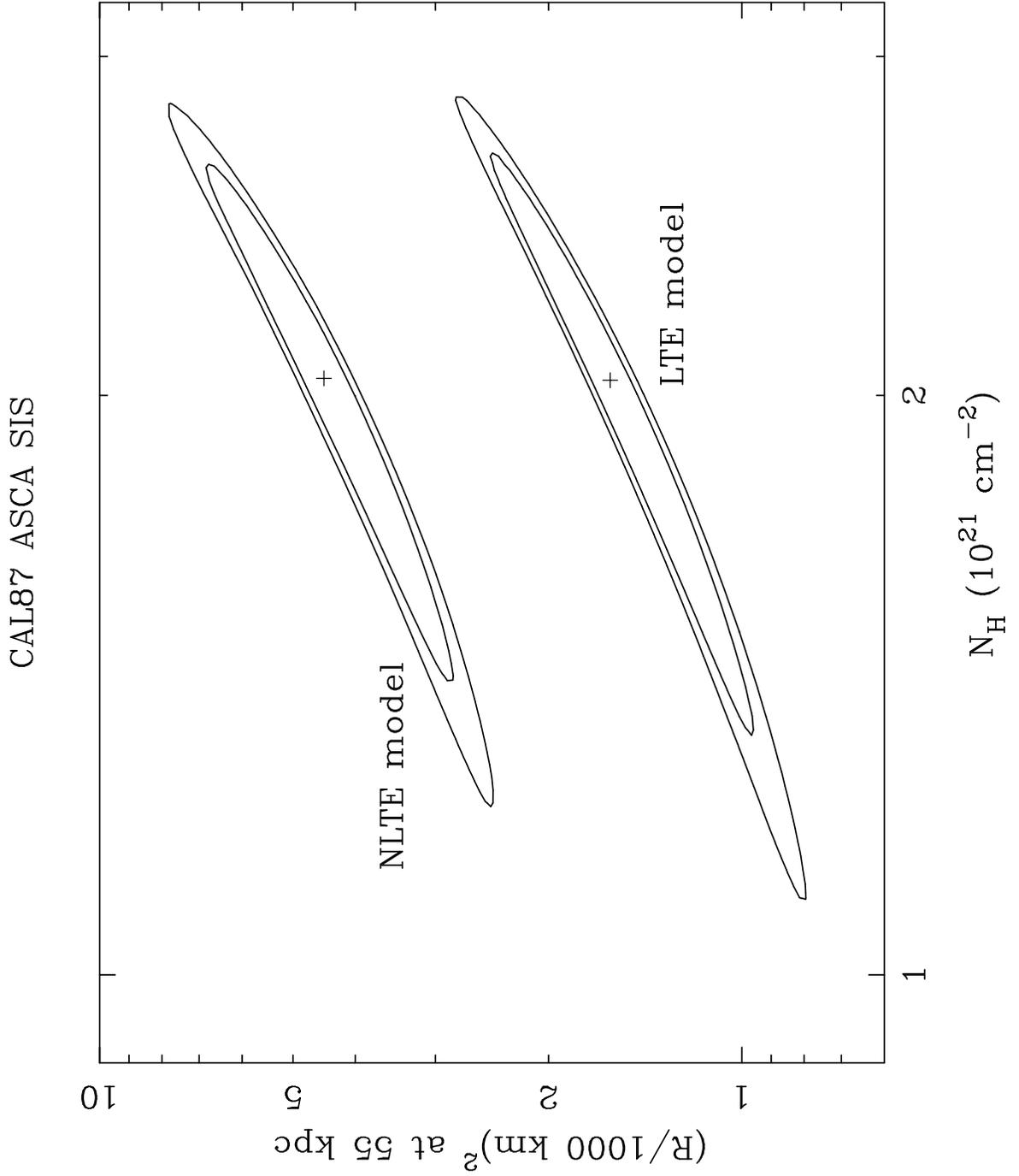}
}
\caption{Confidence limits of the hydrogen column density and the normalization
(converted to the effective emission area  at 55 kpc assuming the 
isotropic emission)
for the CAL87 LTE and NLTE model fits.  The two contours levels indicate the single parameter
90 \% confidence limit ($\chi^2_{min}+2.7$) and the two parameter 90 \% confidence limit 
($\chi^2_{min}+4.6$).}\label{cal87_contour}
\end{figure}

Results of the NLTE model ($g=10^9$ cm s$^{-2}$) fit, which is equally successful,
are shown
in Figure \ref{cal87_nlte_fit} and Table \ref{cal87_nlte_fit_table}.  
A contour map of the hydrogen column density and the
effective emission area are shown
 in Figure \ref{cal87_contour} with that for the LTE model.
Compared to the LTE model, 
the best-fit temperature is significantly lower; this
is considered due to  photoionization effects.
The importance 
of photoionization may be roughly 
estimated using the ionization parameter $\xi \equiv L/nr^2$, 
where $L$, $n$ and  $r$ are luminosity, density and radius, respectively.
On the nuclear-burning white dwarf with a surface gravity of $10^9$ cm s$^{-2}$ and 
an effective temperature of $\sim$ 60 eV,  $\xi$ is estimated as 1 -- 10.
This is large enough to ionize 
C, N, O and Ne to the helium-like or hydrogenic 
stages (e.g., Kallman and McCray 1982).
Therefore, the effects of photoionization are significant, and 
the same ionization degree as 
with the LTE model is achieved at  a lower temperature in the NLTE model
(Hartmann and Heise 1997).
Compared to the LTE model, the hydrogen column density is smaller and the 
best-fit surface area is twice as  large in the NLTE model 
because of the significantly smaller temperature
(Figure \ref{cal87_contour}).

\subsubsection{Effects of Surface Gravity and Elemental Abundances}\label{cal87_gravity_abundance}
The maximum effective temperature expected on the white dwarf surface is 
constrained   by the Eddington luminosity, and may be 
expressed as  $T_{eff}^{(max)} \sim 0.5 \; g^{1/4}$ eV, where $g$ is the surface gravity, 
which is in the range of $\sim 10^7$ -- $10^{10}$ cm $^{-2}$ (Figure \ref{white_dwarf_figure}) and
almost uniquely determined as a function of mass using a standard mass-radius relation
(section \ref{white_dwarf}).
Correspondingly, $T_{eff}^{(max)}$ will be in the range of $\sim$ 30 to 90 eV
depending on  the white dwarf mass.

To see the effect of changing the surface gravity and elemental abundances, 
we calculate theoretical spectral models for 
different surface gravities with the cosmic abundance (metal-richer than the LMC abundance),
and fit the observed spectra.
First, we find that
 the  models with $g=10^8$ cm s$^{-2}$ cannot  account for the
CAL87 energy spectra, since the possible temperatures are 
 too low to produce the observed high energy photons.
Results of the fitting with $g=10^9$  and $10^{10}$ cm s$^{-2}$ are
summarized in Table \ref{cal87_abundance}.

The following points are noticeable from the model fitting results:
(1) Models with different surface gravities can equally fit the data (unless 
surface gravity is too low to attain a sufficiently  hot temperature). 
(2) For the same gravity and  abundances,  always the NLTE model  gives a lower
temperature than the LTE model, because photoionization enables to achieve the same ionization 
state at a lower temperature.  (3) Models with higher gravities require  higher temperatures.
(4) An 
abundance change does not significantly change the results for 
the LTE model, but in the NLTE model
a higher metal abundance gives a slightly higher  temperature.

The point (3) is understood that 
increasing the surface gravity increases the atmospheric density,  and
the higher electron density suppresses  ionization 
(Saha relation;  see Figure \ref{ion_fraction}).  Consequently, 
higher temperature is required to achieve the observed ionization state.
This relation may be written approximately as   $kT_{eff} \propto \log g$.

Effect of the abundance change (point (4) above) may be  understood as 
follows:
Abundance variation  
affects depth of the atmosphere where optical depth reaches unity, such that
increasing metal abundances causes the
absorption edges to be optically thick at closer to the surface where 
temperature is lower.
Hence the number of high energy photons above the edges decreases. 
This effect is more significant in the NLTE model,  since increasing the abundance makes 
the atmosphere more opaque due to  increase of the mass absorption coefficient
(Hartmann and Heise 1997).  Therefore, increasing the metal abundance requires
a higher effective temperature to account for  the observed high energy photons.

\subsection{RXJ0925.7--4758}


\subsubsection{Spectral Analysis with Simple Models}\label{rxj_bb_analsysis}
\begin{figure}
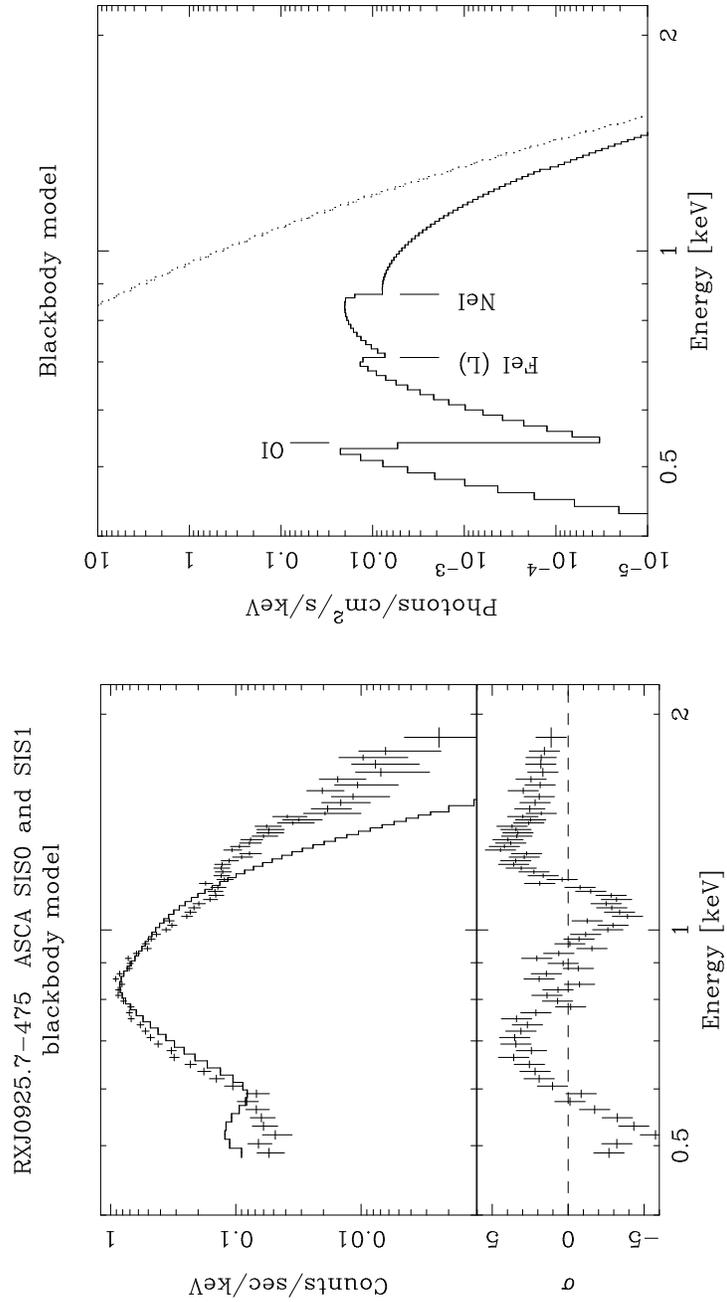

\centerline{
\psfig{figure=f6b.eps,%
      height=9.cm,bbllx=0.cm,bblly=4.cm,bburx=21.cm,bbury=21cm,%
      angle=0,clip=}}
\centerline{
\psfig{figure=f6a.eps,%
      height=9.cm,bbllx=0.cm,bblly=4.cm,bburx=21.cm,bbury=21cm,%
      angle=0,clip=}
}
\caption{Blackbody model fit for RXJ0925.7--4758.
Left: Observed spectrum and best-fit model convolved with the detector response.
The fitting residual is also shown.
Right: The best-fit model before and after  the interstellar absorption.
}\label{rxj_bb_fit}
\end{figure}

First, we show the result of a blackbody plus interstellar absorption model fit
(first column in Table \ref{rxj_bb_table} and  Figure \ref{rxj_bb_fit}).
Although this model does not fit the data at all (reduced $\chi^2$=8), the
following spectral characteristics are clearly recognized: (1) The spectrum is heavily 
absorbed ($N_H \gtrsim 10^{22}$ cm$^{-2}$); 
(2) A deep edge feature such as the \ion{O}{8} edge in CAL87 is {\em not}\/
 observed.
(3) The energy spectrum is much harder than that of CAL87 and extends to $\sim$ 1.5  keV.
A  thin thermal plasma model (Raymond and Smith 1977)
gives $T$= 50 eV, $N_H=1.98 \times 10^{22}$ cm$^{-2}$, and a reduced $\chi^2 $ of 16.
The high absorption and the significant excess above $\sim$ 1.5 keV 
are  model independent.  A  power-law plus absorption model does not
give a meaningful result, ending up with a photon index higher than 10.

\begin{figure}
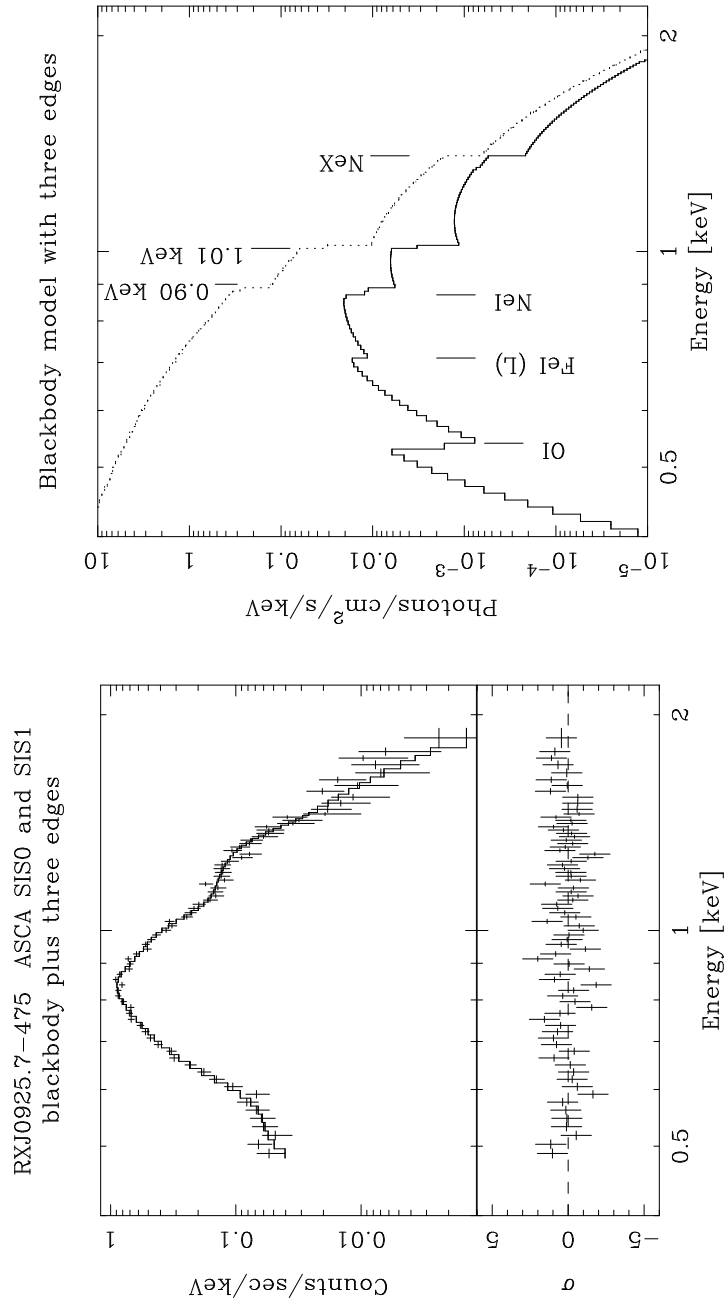

\centerline{
\psfig{figure=f7b.eps,%
      height=9.cm,bbllx=0.cm,bblly=4.cm,bburx=21.cm,bbury=21cm,%
      angle=0,clip=}}
\centerline{
\psfig{figure=f7a.eps,%
      height=9.cm,bbllx=0.cm,bblly=4.cm,bburx=21.cm,bbury=21cm,%
      angle=0,clip=}
}
\caption{Blackbody plus edge model fit for RXJ0925.7--4758.
Three edges are included at 0.89 keV, 1.01 keV and 1.36 keV.
The oxygen abundance in the interstellar absorption is reduced than the
cosmic abundance (see the text).
Left: Observed spectrum and best-fit model convolved with the detector response.
The fitting residual is also shown.
Right: The best-fit model before and after  the interstellar absorption.
}\label{rxj_bb_edge_fit}
\end{figure}

We next try a blackbody model with absorption edges.  We find that three absorption
edges are needed to  fit the energy spectrum (Table \ref{rxj_bb_table},
second column).
In addition, if we allow oxygen abundance in the interstellar medium to be free, 
the fit significantly improves as $\Delta \chi^2 = 53.6$
(Table \ref{rxj_bb_table}, third column).  
Figure \ref{rxj_bb_edge_fit}
shows the satisfactory fit by the blackbody plus three edge model 
with a reduced oxygen abundance. 
The first edge 
at  $0.90\pm^{0.02}_{0.03}$ keV  may be identified as 
 \ion{O}{8}   or \ion{Ne}{1} K-edges  at 0.87 keV.
Instead of putting a 0.90 keV edge, the spectrum can be fitted equally well
by increasing the neon abundance in the interstellar absorption (Table \ref{rxj_bb_table},
fourth column).
The highest edge at  $1.36\pm^{0.07}_{0.05}$ keV
could be identified with the \ion{Ne}{10} edge at 1.36 keV.  
The   $1.02\pm0.02$ keV edge, whose origin is unclear,
is close to the energies of
the \ion{Fe}{7} L$_{\rm I}$-edge (1.03 keV) or
 \ion{Ne}{6} K-edge (1.04 keV).



The hydrogen column densities ($\sim 1-2 \times  10^{22}$
 cm$^{-2}$) 
are consistent with those estimated from optical/IR  and ROSAT observations
(Motch, Hasinger and Pietsch 1994).  The  large column density suggests the source
is beyond the Vela Sheet molecular cloud at 425 pc (Motch, Hasinger and Pietsch 1994).


\subsubsection{Theoretical White-dwarf Models}
\label{rxj_gravity_abundance}
\begin{figure}
\centerline{
\psfig{figure=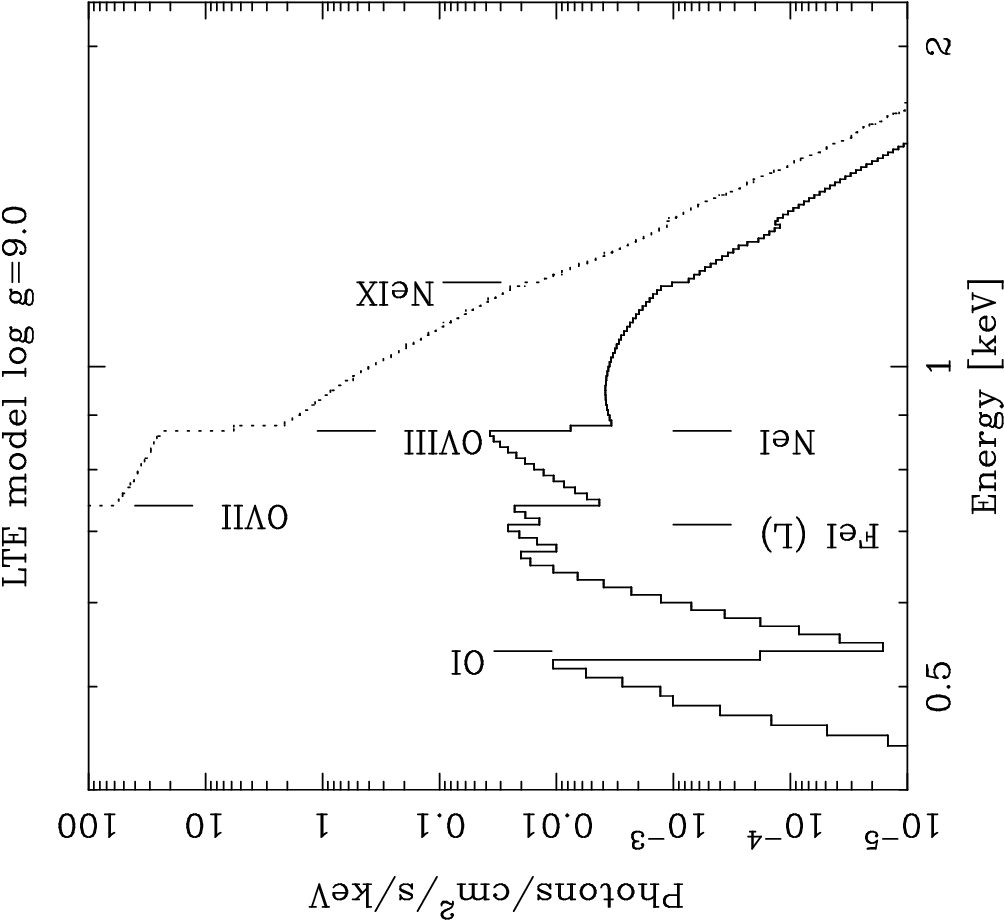,%
      height=9.cm,bbllx=0.cm,bblly=4.cm,bburx=21.cm,bbury=21cm,%
      angle=0,clip=}}
\centerline{
\psfig{figure=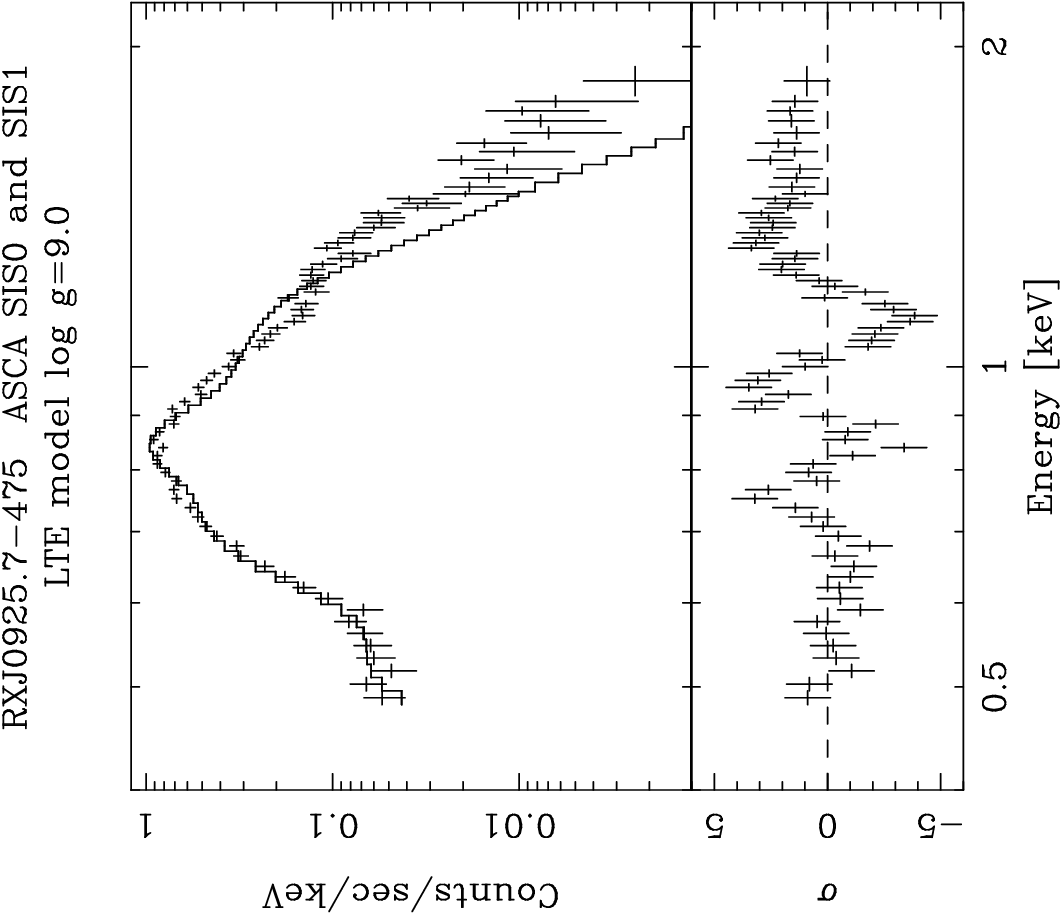,%
      height=9.cm,bbllx=0.cm,bblly=4.cm,bburx=21.cm,bbury=21cm,%
      angle=0,clip=}
}
\caption{LTE model fit for RXJ0925.7--4758 with the surface gravity
 $10^{9}$ cm s$^{-2}$.
The oxygen and neon abundances in the interstellar absorption are made
free parameters  (see the text).
Left: Observed spectrum and best-fit model convolved with the detector response.
The fitting residual is also shown.
Right: The best-fit model before and after  the interstellar absorption.
}\label{rxj_lte_log_9}
\end{figure}

\begin{figure}
\centerline{
\psfig{figure=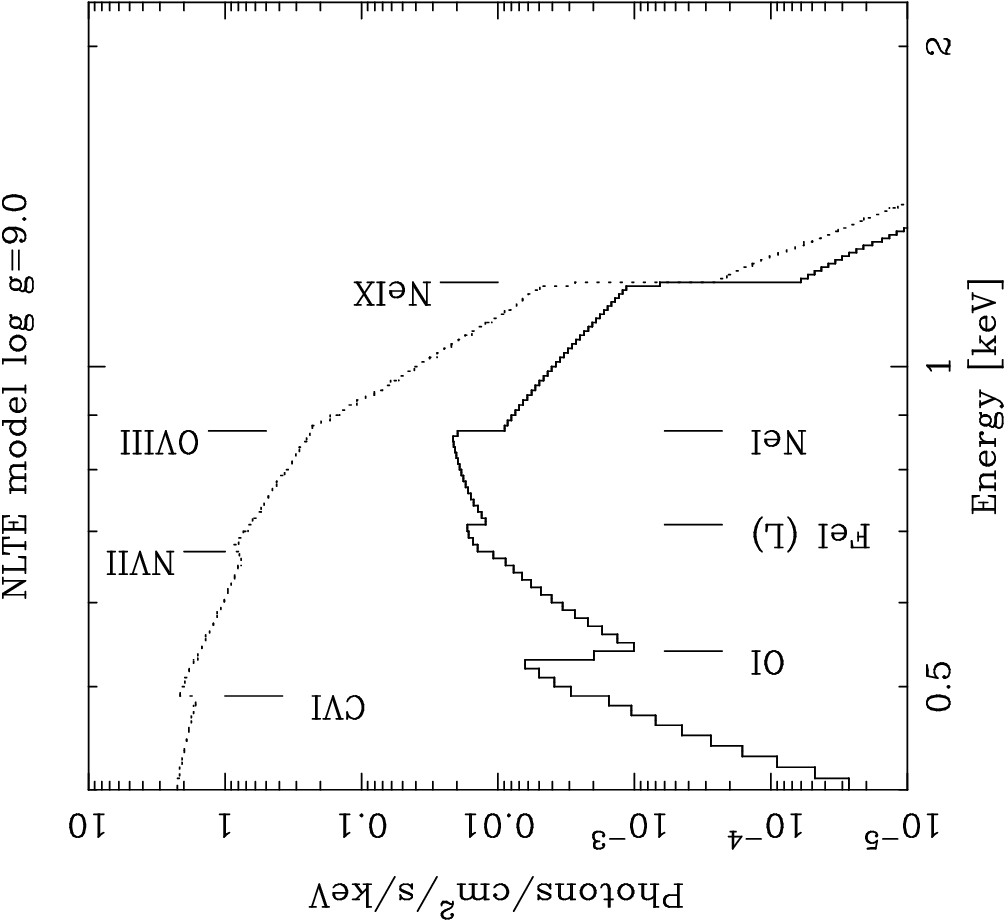,%
      height=9.cm,bbllx=0.cm,bblly=4.cm,bburx=21.cm,bbury=21cm,%
      angle=0,clip=}}
\centerline{
\psfig{figure=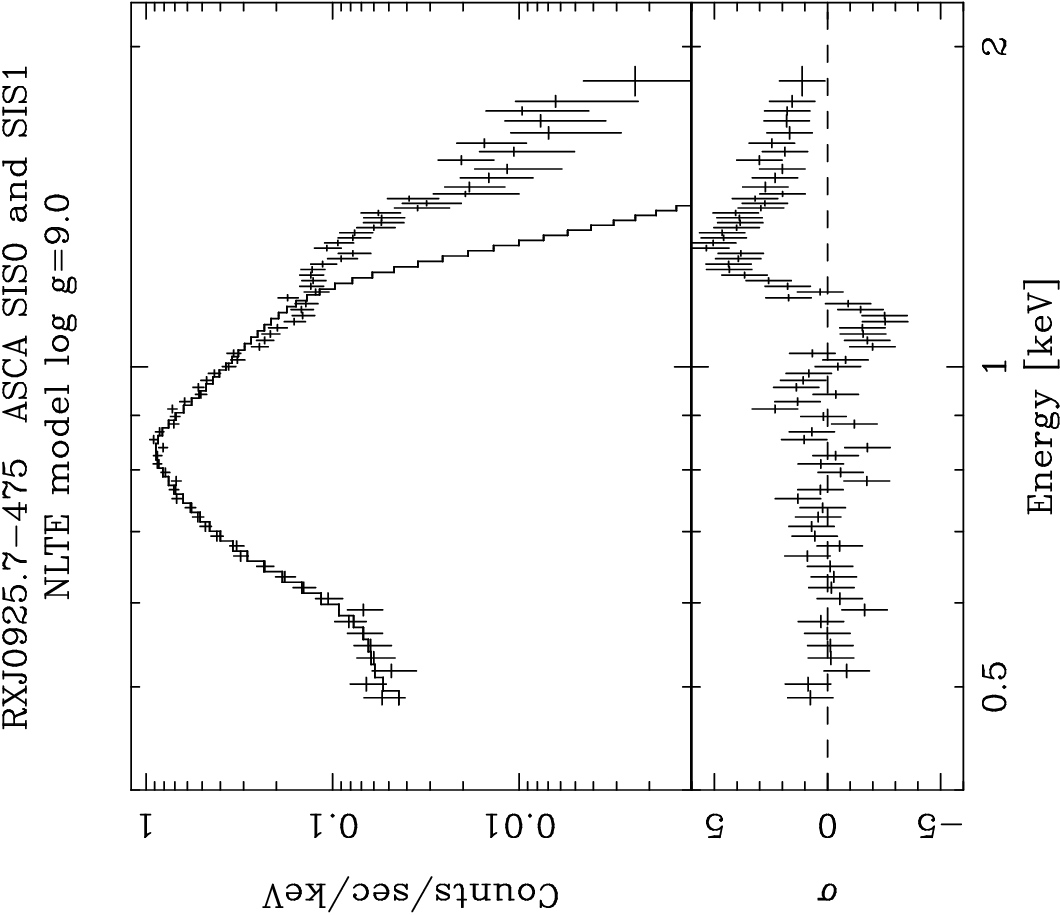,%
      height=9.cm,bbllx=0.cm,bblly=4.cm,bburx=21.cm,bbury=21cm,%
      angle=0,clip=}
}
\caption{NLTE model fit for RXJ0925.7--4758 with the surface gravity
 $10^{9}$ cm s$^{-2}$ and the cosmic abundance.
The oxygen and neon abundances in the interstellar absorption are made
free parameters  (see the text).
Left: Observed spectrum and best-fit model convolved with the detector response.
The fitting residual is also shown.
Right: The best-fit model before and after  the interstellar absorption.
}\label{rxj_nlte_log_9}
\end{figure}

\begin{figure}
\centerline{
\psfig{figure=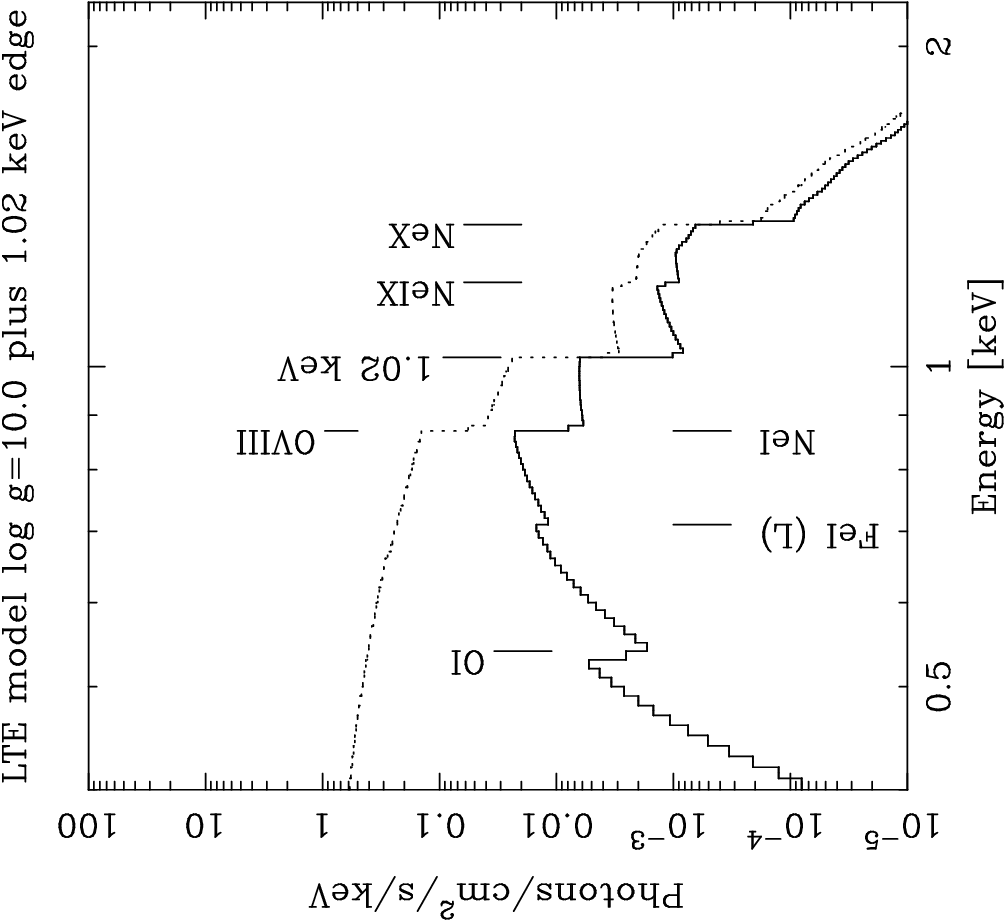,%
      height=9.cm,bbllx=0.cm,bblly=4.cm,bburx=21.cm,bbury=21cm,%
      angle=0,clip=}}
\centerline{
\psfig{figure=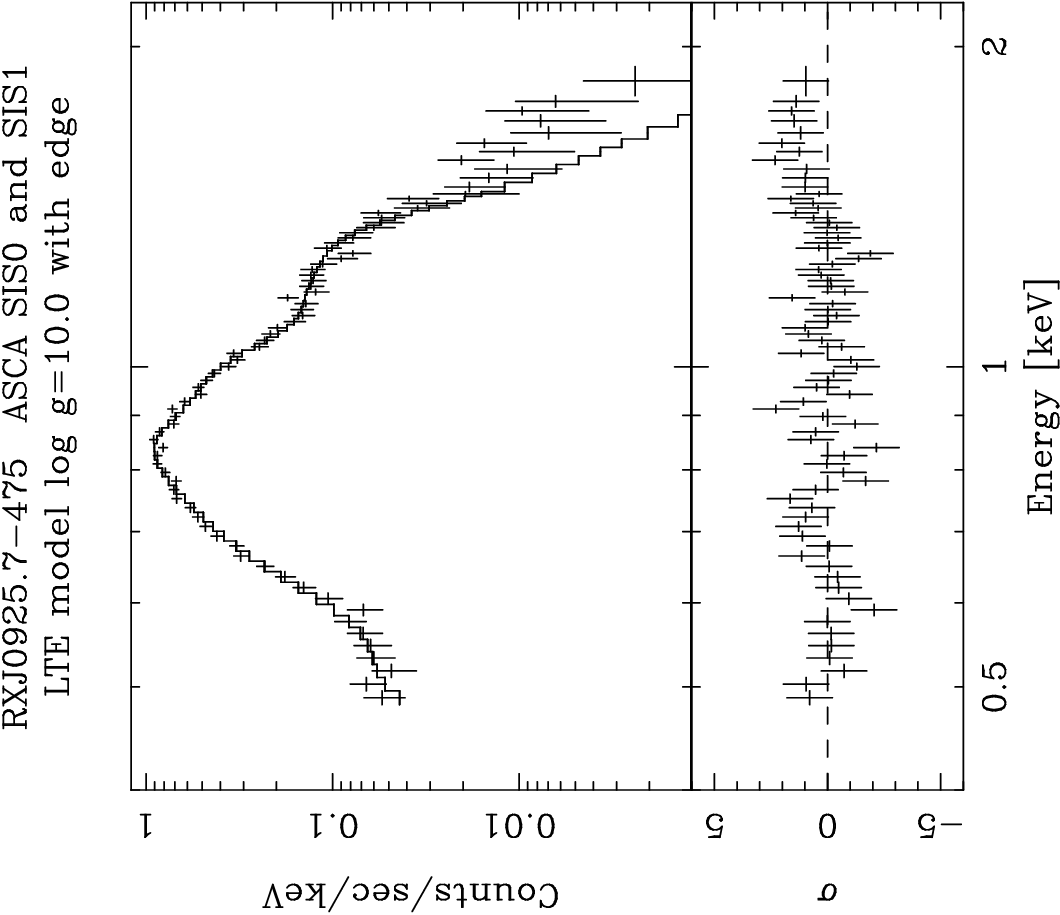,%
      height=9.cm,bbllx=0.cm,bblly=4.cm,bburx=21.cm,bbury=21cm,%
      angle=0,clip=}
}
\caption{LTE model fit for RXJ0925.7--4758 with the surface gravity
 $10^{10}$ cm s$^{-2}$ and an absorption edge at 1.02 keV.
The oxygen and neon abundances in the interstellar absorption are made
free parameters  (see the text).
Left: Observed spectrum and best-fit model convolved with the detector response.
The fitting residual is also shown.
Right: The best-fit model before and after  the interstellar absorption.
}\label{rxj_lte_log_10_edge}
\end{figure}

Next, we try LTE and NLTE white dwarf spectral models (cosmic abundance of the
white dwarf atmosphere is assumed). 
The oxygen and neon abundances in the interstellar absorption are allowed 
to be free, 
which will be reasonable from the consideration in the previous section.  
Results are shown in Table \ref{rxj_lte_nlte_table} and Figures \ref{rxj_lte_log_9} to 
\ref{rxj_nlte_log_10_edge}.

Figures \ref{rxj_lte_log_9} and \ref{rxj_nlte_log_9} respectively show examples of the fits  with 
the LTE and NLTE model
with a surface gravity of $10^9$ cm s$^{-2}$.
These model spectra have very strong \ion{O}{8} edge at 0.89 keV 
(LTE model) or \ion{Ne}{9} edge at 1.20 keV (NLTE model), but neither
of them is actually observed. In fact,
there are significant amounts of hard photons above these 
edge energies, and the observed spectrum indicates the  \ion{Ne}{10} edge 
at 1.36 keV,  instead of the \ion{Ne}{9} edge in the NLTE model.
A more sophisticated NLTE model including metal line opacities shows 
 weaker \ion{Ne}{9} edge but more conspicuous
\ion{O}{8} edge   (Hartmann et al.\ 1999), 
rather being similar to the LTE model.  
Hence, the problem that there are much more 
high energy photons than predicted by our LTE or NLTE model
is unlikely to be solved by further minor improvement of the models.

At the surface gravity  $10^9$ cm s$^{-2}$, the maximum effective temperature
will be $\sim 90$ eV (section \ref{cal87_gravity_abundance}).  On the other hand, 
the effective temperature has to be as high as  $\sim$ 100 eV so that 
\ion{Ne}{10} becomes  more dominant than  \ion{Ne}{9} (Figure \ref{ion_fraction}).
Therefore, 
expecting a higher temperature and ionization state, 
we next try models with the surface gravity $10^{10}$ cm s$^{-2}$, which 
would be considered the upper-limit for 
white dwarfs (Figure \ref{white_dwarf_figure}).  
The temperature grids are every  $ 10^5$ K 
from 8 $\times 10^5$ K to $1.9 \times 10^6$ K 
for the LTE model and  every  $10^4$ K from 8 $\times 10^5$ K to $1.2 \times 
10^6$ K for the NLTE model.
Although models do exhibit \ion{Ne}{10} edge,
the fit does not improve significantly (Table \ref{rxj_lte_nlte_table})
 unless putting an additional edge at $\sim 1.02 $ keV.
In Figure \ref{rxj_lte_log_10_edge}, \ref{rxj_nlte_log_10_edge} 
and Table \ref{rxj_lte_nlte_table}, we show results of the successful LTE and NLTE model fits with the surface
gravity $10^{10}$ cm s$^{-2}$ and an absorption edge at
$\sim 1.02 $ keV.  

If the CNO cycle is dominant on the white dwarf surface, it
will change the relative abundances of these elements, such 
that C and O abundances are suppressed relative to that of N.
Having this in mind, we have calculated several LTE and NLTE
models with non-standard element abundances.  However,
we could not fit the RXJ0925.7--4758 spectrum successfully, unless
putting the $\sim 1.02 $ keV edge. For example, we have tried a LTE model
with [N/C]=200 and [N/O] = 50. The \ion{O}{7} and \ion{O}{8}
edges become weak, which is in the sense to fit the data, but
the \ion{N}{6} (0.55 keV) or \ion{N}{7} (0.67 keV)  edges 
are very conspicuous in the model, which are not observed.
We have also tried models in which all the heavy element
abundances are lowered by an order of magnitude.  We have found 
that reducing the abundance helps to account for  the hard-tail 
above the \ion{Ne}{9} edge, which is in agreement  with 
Shimura (2000).
However,  the 1.02 keV edge feature still remains in the residual.

We have also tried a two component model, expecting that the hard photons 
above \ion{O}{8} and  \ion{Ne}{9} edges might   be
explained by an additional hard component.  
Hartmann et al.\ (1999) fitted the RXJ0925.7--4758 spectrum 
observed with  LECS on-board SAX with a NLTE model for
the soft component  and a thin thermal model for the hard component.
We have tried LTE model with $g= 10^{10}$ cm s$^{-2}$ with an additional 
thin thermal plasma model (Raymond and Smith 1977) for the hard component, 
according to Hartmann et al.\ (1999).  The results are shown in Table \ref{rxj_lte_ray_table}
and Figure \ref{rxj_lte_ray}.  Although the high energy photons are explained
by the hard component as expected, the fit is not as good as
the previous ones including  the $\sim$ 1.02 keV  edge.
In fact,   we still see the $\sim$ 1.02 keV edge feature clearly
in the residual
of the fit (Figure \ref{rxj_lte_ray}).
Note that ASCA SIS has about twice better spectral resolution than
SAX LECS at 1 keV, and residual of the SAX two 
component fit (Figure 7 in Hartmann et al.\ 1999) also indicates
an edge-like feature at $\sim 1 $ keV.
Therefore, we conclude that 
the $\sim$ 1.02 keV edge is a local feature, and cannot
be explained by introducing additional emission component.


\section{Discussion}

\subsection{CAL87}

\subsubsection{White dwarf parameters}\label{white_dwarf}

We have applied theoretical LTE and NLTE 
models to the CAL87 energy spectra observed with ASCA,
and obtained effective temperature and emission area.
Based on the spectral fitting results, we can constrain white dwarf 
parameters in CAL87.

Mass and radius of white dwarfs are related by the theoretical white-dwarf
mass-radius relation, thus radius and surface gravity are 
almost uniquely determined as a function of mass.  In the
bottom panel of Figure \ref{white_dwarf_figure}, we show
white dwarf radius and surface gravity as a function of mass, 
in which we used an approximated formula of the mass-radius relation
by Pringle and Webbink (1975). Although the mass-radius relationship
is known to be dependent on internal atomic compositions
(e.g., Panei, Althaus and  Benvenuto 2000), this uncertainty hardly
affects our results, since most uncertainties in our discussion
originate in the systematic difference of the LTE and NLTE models and
statistical errors from  the spectral fitting.

The Eddington luminosity is a function of mass, hence 
the corresponding maximum effective temperature on the white dwarf surface
is also determined solely by mass (section \ref{cal87_gravity_abundance});
these are shown in the top and middle panel of Figure \ref{white_dwarf_figure}
respectively.  

From model fitting  assuming  the surface gravity $g = 10^9$ cm s$^{-2}$, 
we have obtained effective temperatures.  
The model spectral shape is hardly
dependent   on the surface gravity, and best-fit effective temperature 
varies with the gravity as $kT_{eff} \propto \log g$ (section \ref{cal87_gravity_abundance}).
This relation is shown for the LTE and NLTE fitting results 
respectively in the middle panel of
Figure \ref{white_dwarf_figure} by two solid lines.  
The total bolometric luminosity is calculated as
$4 \pi \sigma R^2 T_{eff}^4$, using the observed
effective temperature and the white dwarf radius 
derived from the mass-radius relationship,
which is also indicated in the top panel
by two solid lines for LTE and NLTE models respectively.
Note that the intrinsic luminosities thus
calculated are much larger than the apparent observed luminosity derived 
from the observed flux
assuming uniform emission, which suggests most of the
emission is blocked  (see below).

Steady nuclear burning of the hydrogen accreted onto white dwarfs,
to be  observed as super-soft X-ray sources, 
takes place only in a limited range of the mass accretion rate,
(1 -- 5)$ \times 10^{-7} M_\odot $yr$^{-1}$ (van den Heuvel et al.\ 1992).  
The allowable luminosity range with the steady nuclear burning calculated  by
van den Heuvel et al.\ (1992)  is shown as shaded area
in the top panel of Figure \ref{white_dwarf_figure}.
In this panel, portion of the two lines marked LTE and NLTE 
lying on the allowable theoretical 
luminosity range should give
the realistic luminosity and mass ranges.
They are, $1.0-1.2 M_\odot$ and $6\times10^{37}-1.2\times10^{38}$ ergs s$^{-1}$
for the LTE model, and  
$0.8-1.0 M_\odot$ and $(4-7)\times10^{37}$ ergs s$^{-1}$
for the NLTE model, respectively.
Considering current uncertainties of theoretical 
spectral models, we conclude the total bolometric
luminosity and mass of CAL87 are in the range of
$4\times10^{37}-1.2\times10^{38}$ ergs s$^{-1}$ and
$0.8-1.2 M_\odot$, respectively.

In the top panel of Figure \ref{white_dwarf_figure},
we also  show the range of the  bolometric luminosity
calculated from the observed flux assuming uniform emission.
We can see that the
observed luminosity, $4\times10^{36}$ to $10^{37}$ erg s$^{-1}$, is
about an order of magnitude smaller than the expected  luminosity, 
indicating that only $\sim 10 $ \% of the total emitted X-rays reaches us.
This is consistent with, and strengthens the idea by Schmidtke et al.\ (1993)
and Schandl, Meyer-Hofmeister and Meyer (1997) that the white dwarf in CAL87
is permanently blocked from line-of-sight by an outer part of the accretion disk, and 
we are observing scattering emission by the Accretion Disk Corona
(ADC).
The scattering optical depth of the corona is estimated as $\sim$ 0.1, 
which is small enough not to smear the observed deep edge feature
(section \ref{cal87_simple}). 

Hydrogen column density of the ADC will be $\sim 1.5 \times 10^{23}$ cm$^{-2}$, 
and its size  is estimated as $\sim 5 \times 10^{10}$cm
(see next section).  Correspondingly, the ionization parameter, $\xi = L/(nr^2) \approx L/(N_H r)$
will be  5000 -- 16000.  This will be large enough to  completely photoionize
heavy elements besides iron, so that absorption edges  are fully eliminated below
$\sim 8 $keV (e.g., Kallman and McCray 1982).  
Hence, the ADC should be  observed only as a scatterer, 
not an absorber, which is consistent with the present observation
(see also discussion in Asai et al.\ 1998).

\subsubsection{Orbital Eclipses}\label{cal87_eclipse}

Based on the white dwarf parameters obtained  in the previous section,
we will try to explain the  orbital eclipse in the framework
of the ADC model.  Namely,  a part of the extended emission 
from ADC is blocked by the companion star, resulting a shallow
and extended X-ray eclipse as observed 
(Schmidtke et al.\ 1993; Schandl, Meyer-Hofmeister and Meyer 1997).
We will carry out Monte Carlo simulation searching for the orbital parameters
and ADC configuration to reproduce the observed X-ray orbital light  curve.

In our model, parameters which affects the orbital light curve
are the following:  the white dwarf mass ($M_{WD}$), companion mass
($M_{C}$), size of the ADC,  and the orbital inclination angle ($i$).
In the previous section we estimated $M_{WD}$ as 0.8--1.2
$M_\odot$.  In the simulation, we assume $M_{WD}=1 M_\odot$.
 $M_{C}$ is expected to be in the range of 1.4 $M_\odot$
to  2.2 $M_\odot$ from the  binary evolution requirements
(van den Heuvel et al.\ 1992); we assume $M_{C} = 1.5 M_\odot$.
The orbital period is 10.6 hr (Alcock et al.\ 1997 and references therein), 
and the binary separation is
determined from the Kepler's law as $a = 2.3 \times 10^{11}$ cm. 
The Roche lobe radius is calculated using the formula by Pringle (1985),
and we consider that the companion is filling its  Roche lobe. 
Roche radius for the companion and white dwarf will be 
0.41 times and 0.34 times the binary separation, respectively.  Accretion
disk radius is assumed to be 0.8 times the white dwarf Roche lobe
radius, following
Schandl, Meyer-Hofmeister and Meyer (1997).
The ADC is assumed to be a sphere whose radius is smaller than the
accretion disk radius.  The white dwarf is permanently blocked from the
line-of-sight by the outer flared parts of the disk or the ``spray''
region (Schandl, Meyer-Hofmeister and Meyer 1997).  This is approximated
by
 that the disk is a slab having a constant thickness 0.2 times the disk radius.
Also disk is assumed not to be tilted from the orbital plane.

We have 
carried out 
Monte Carlo simulations for many different combinations of these parameter values,
and found that the orbital light curve is most strongly dependent on the  orbital 
inclination,  and secondly on the  ADC size.  
In particular, we have found that depth of the eclipse is very sensitive
to the inclination, since in our  model a tiny  portion of the ADC 
has to be seen just behind the companion (see the diagram at $\phi = 1$
in Figure \ref{light_curve1}).
Dependence on other parameters is relatively minor.  
In any case, it is impossible to constrain
orbital parameters from our simulation, since there are too many combinations which
can reproduce the observed X-ray light curve, which is rather noisy.

As a typical
 example of successful orbital simulation, we show the result with
the ADC size $4.8 \times 10^{10}$ cm (= 0.75 times the accretion disk radius), 
and $i=73^\circ$ (Figure \ref{light_curve1}).
Schandl, Meyer-Hofmeister and Meyer (1997) carried out orbital simulation,
and successfully  explained the observed  optical light curve with
 $i=77^\circ$.
We have reached $i=73^\circ$, 
having searched for the best inclination angle around $i=77^\circ$
to  fit the observed X-ray light curve.

Several  remarks should be made here on our simulation:
If we compare closely, the  observed eclipse is
slightly wider than the simulation (Figure \ref{light_curve1}).
We assumed that the ADC has constant emissivity and did not take
into account the rim darkening,  inclusion of which
should further narrow the simulated eclipse.  
Our choice of the ADC size (0.75 times the disk radius) is rather arbitrary,
but if the ADC size becomes  smaller than half the disk radius, 
the eclipse will be still narrower and
very fine tuning of the inclination will be required so that the
eclipse takes place.
Hence, our model favors a large ADC size. Also
there expects to be additional mechanisms to widen the eclipse
which we have not taken into account. 
For example, 
we assumed that the companion is a perfect sphere with definite  boundaries, but in reality the companion filling the Roche lobe should be  elongated toward the white dwarf, which
can  block the ADC more effectively and make the eclipse wider.
In addition, the companion may have an extended atmosphere and/or winds to attenuate
X-rays by scattering, which also will work to widen
the eclipse  (Asai et al.\ 1998).  Future high throughput observations 
will allow more precise 
X-ray eclipse analysis to constrain the binary parameters tightly.

\subsection{RXJ0925.7--4758}\label{rxj_discussion}

We have found that the RXJ0925.7--4758 spectrum has high energy photons
above \ion{O}{8} or \ion{Ne}{9} edge energies and exhibits 
the \ion{Ne}{10} edge at 1.36 keV.  Such a hot atmosphere, 
as high as $T_{eff} \sim $ 100 eV (Tables \ref{rxj_lte_nlte_table}), 
requires an extremely high surface gravity,  $g \approx 10^{10}$ cm s$^{-2}$,
which may be achieved only at close to the Chandrasekhar limit, $\sim 1.4 M_\odot$
(Figure \ref{white_dwarf_figure}; see also Hoshi 1998;   Shimura 2000).
Presence of such an extremely massive white dwarf is 
also consistent with  optical observations
of the binary motion.
(Schmidtke et al.\ 2000).

We have found that the model fits always require
an additional absorption edge at  $\sim1.02 $ keV.
This edge feature does not disappear by introducing second  high
energy spectral component (Figure \ref{rxj_lte_ray}), which was
proposed 
by Hartmann et al.\ (1999) to explain the SAX RXJ0925.7--4758 spectrum.

Near the Chandrasekhar limit, the X-ray luminosity 
expected from the  steady nuclear burning will be as high as (1--2)$\times 10^{38}$erg s$^{-1}$
(Figure \ref{white_dwarf_figure}).  On the other hand,
the luminosity we obtained from the LTE or NLTE model fitting 
($g \approx 10^{10}$ cm s$^{-2}$, including the 1.02 keV edge)
is (3--4)$\times 10^{34}$erg s$^{-1}$ at 1 kpc.
Even if we put the source  at $\sim 10 $ kpc,  discrepancy 
between the expected and observed  luminosities
 is still almost two orders or magnitude.
We have found that 
the observed luminosity of CAL87 is  just 10 \% of the 
expected  total luminosity (section \ref{white_dwarf}), which 
we consider  because the white dwarf is always hidden
by the accretion disk.  Blocking by the accretion disk
will be unlikely
for RXJ0925.7--4758 though, since neither X-ray or optical eclipses
has been  detected, and the orbital inclination is 
considered to be much smaller (Motch 1998; Schmidtke 2000).

To solve the luminosity discrepancy, we propose a model that 
the white dwarf in RXJ0925.7--4758 is behind a non-uniform,
almost fully ionized Thomson thick cloud,  so that heavy
electron scattering causes the significant  luminosity reduction.  
Since the cloud is not spherical,  photons 
scattered out of the line of sight will never be  observed again.
The 1.02 keV absorption edge and the strong low energy absorption 
may be explained simultaneously, if 
ionization state of the cloud in the line of
sight is not uniform, 
so that less ionized part of the cloud is responsible
for these absorptions.
A transient optical jet has been observed from 
RXJ0925.7--4758  (Motch 1998).  Therefore, we consider it
likely that
matter previously emitted as the jet from the binary system 
lies in the line of sight to scatter the X-rays from the white dwarf.

We consider the observed luminosity, $\sim 3 \times 10^{34} 
(d/1\; {\rm kpc})^2$, is $ e^{-\tau_{sct}}$ times the
actual luminosity, $\sim 10^{38} $ erg s$^{-1}$, where $\tau_{sct}$
is the scattering optical depth of the intervening matter.
Hence, column density of the intervening matter may be written as,
\begin{equation}
N_H \sim \{12 - 7 \log (d/1 \;{\rm kpc})\} \times 10^{24} {\rm cm}^{-2}.
\end{equation}
On the other hand, time average of the mass ejection rate due to the
jet is,
\begin{equation}
        <dM/dt> = 4 \pi r^2 v n (\delta/4 \pi) w m_H,
\end{equation}
where $v$ is the jet velocity, $n$  the number density,
$w$  the time fraction when the jet takes place, 
$\delta$ solid angle, and  $m_H$ is the mass per hydrogen atom.
The column density can be also written as,
\begin{equation}
N_H = \int_R^\infty n w dr = \frac{<dM/dt>}{R v \delta m_H},
\end{equation}
where  $R $ is the white dwarf radius.
From these relationships, we  get,
\begin{equation}
<dM/dt> \sim 3 \times 10^{17} \left(\frac{R}{3000 \; {\rm km}}\right)
\left(\frac{v}{5000\; {\rm km/s}}\right)
\left(\frac{\delta}{1}\right)
\left\{12-7\log\left(d/1\; {\rm kpc}\right)\right\} \;{\rm g\, s^{-1}},\label{mdot}
\end{equation}
where plausible values of $v$ and $\delta$ are taken
from Motch (1998).

Only constraint we may impose is that the mass ejection rate
does not exceed the plausible mass accretion rates
for the super-soft sources,  $\sim 10^{19}$ g\, s$^{-1}$
(e.g., van den Heuvel et al.\ 1992).
From equation \ref{mdot}, we see that this condition is satisfied
as long as  $d \gtrsim$ 1 pc; hence we may not constrain the distance
to the source.
For an assumed  distance of 1 kpc, 3 kpc and 10 kpc,
the column density of the intervening matter necessary to
reduce the luminosity  will be
$\sim 1 \times 10^{25}$, $\sim 9\times 10^{24}$, 
$\sim 5 \times 10^{24}$ cm$^{-2}$ respectively, and the mass
ejection rate is $\sim 4 \times 10^{18}$, 
$\sim 3 \times 10^{18}$ and $\sim 2 \times 10^{18}$ g\,s$^{-1}$.

Ionization parameter of the matter may be written as,
$$
\xi =\frac{L}{nr^2} = \frac{Lw}{N_H R}
$$
\begin{equation}
\sim 1200 \frac{(L/10^{38} \;{\rm erg \;s^{-1}})(w/0.04)}
{(N_H/10^{25}\;{\rm cm}^{-2})(R/3000 \;{\rm km})},
\end{equation}
where we took the jet frequency $w \sim 0.04 $, as the jet was observed
in one out of 23 nights (Motch 1998). 
The plasma is almost fully ionized, that it is transparent 
for low energy absorption, but works just as a scatterer
to reduce the observed luminosity.  The jet frequency
may not be uniform, and there may  be a period when 
$w$ is small and a particularly dense material is ejected.
In that case, the radial region corresponding that
dense matter will have  locally a lower ionization state.
The 1.02 keV edge may originate in 
moderately ionized heavy elements such as Ne  or
Fe in such a low-ionized region.

\section{Conclusion}

We have observed the super soft sources CAL87 and RXJ0925.7--4758  with 
ASCA SIS, and carried out precise spectral analysis.  
Thanks to the superior energy resolution of SIS ($\Delta E/E \sim 10 \% $
at 1 keV) compared to previous instruments,  we could study
detailed X-ray spectral structure of these  sources for the first time.
Important results are summarized
in the following:

(1) We have applied theoretical spectral models to CAL87, and
constrained the white dwarf mass and intrinsic luminosity as 
$0.8 - 1.2 M_\odot $ and $4 \times 10^{37}$--$1.2 \times 10^{38}$
erg s$^{-1}$, respectively.  We have found the observed luminosity
is an order of magnitude smaller than the intrinsic luminosity, 
which indicates the white
dwarf is permanently blocked by the accretion disk and not directly seen.
This strongly suggests that  we 
are observing a scattering emission by a fully ionized 
accretion disk corona (ADC) whose column density is $\sim 1.5 \times 10^{23}$ cm$^{-2}$.

(2) Through simulation, we have shown that the orbital eclipse can be explained 
by the ADC model, such that a part of the extended X-ray emission from the ADC,
whose size is expected to be  $\sim$ 75 \% of the accretion disk radius, 
is blocked by the companion star filling its Roche lobe.
The orbital inclination angle is  $\sim 73^\circ$ to explain
the observed eclipse profile.

(3) In order to explain the RXJ0925.7--4758 spectrum, 
 very high surface gravity and temperature, $\sim 10^{10}$
 cm s$^{-2}$ and $\sim $100 eV respectively,  are required. 
These values are  only possible for an extremely heavy white dwarf near the Chandrasekhar
limit.  Therefore, RXJ0925.7--4758
may be an immediate progenitor of Type Ia supernovae, 
which has been  long-sought (e.g., Livio et al.\ 1996).
We have found the energy spectrum has a $\sim$ 1.02 keV edge, which may be
a composite of mildly ionized L- or K-edges of
heavy elements.

(4) Although  super-soft sources should have the  luminosity 
 $\sim 10^{38}$ erg s$^{-1}$ at the Chandrasekhar limit, 
the observed luminosity from RXJ0925.7--4758 is smaller by more than
two orders of magnitude.
To explain the luminosity discrepancy,  
we propose a model that very thick 
matter (as much as  $\sim 10^{25}$ cm$^{-2}$)
which was
previously ejected from the system as jets
intervenes  the line of sight, and reduces the luminosity significantly
due to Thomson scattering.  We have shown that the matter is almost
fully ionized that it is transparent for soft X-ray photons.
Some  radial portion of the ejected matter,
 where ionization state is locally
low and heavy elements are not fully ionized,  
may be origin of the significant  low energy absorption and
the  $\sim$ 1.02 keV edge.

We wish to thank the anonymous referee for his/her
carefully reading the manuscript and 
useful comments.




\clearpage
\begin{table}
\begin{center}
\caption{CAL87 blackbody plus edges model fit for the average spectrum}\label{bb_edge_table}\label{bb_table}
\begin{tabular}{lcc}\hline\hline
$N_H [10^{21}$cm$^{-2 }]$            & 7.7$\pm^{5.4}_{4.3}$ &3.8$\pm$0.5 \\
$T$ [eV]                             & 53$\pm^{26}_{14}$    & 75 (fixed)  \\
Edge1$\left\{ \begin{array}{l}
        E_{edge} \; [{\rm keV}]\\ 
        \tau
      \end{array}\right.$            &   $\left.\begin{array}{l}
                                       0.739 \; ({\rm fixed}) \\ 
                                       0.49\pm0.37\\
                                      \end{array}\right.$ &   $\left.\begin{array}{l}
                                                           0.739 \; ({\rm fixed}) \\ 
                                                           0.73\pm0.30 \\
                                                           \end{array}\right\}$ 
                           \\
Edge2$\left\{ \begin{array}{l}
        E_{edge} \; [{\rm keV}]\\ 
        \tau
      \end{array}\right.$            &  $\left.\begin{array}{l}
                                       0.871 \; ({\rm fixed}) \\ 
                                       10.0 \; ({\rm fixed})\\
                                        \end{array}\right.$   & $\left.\begin{array}{l}
                                                               0.871 \; ({\rm fixed}) \\ 
                                                              10.0 \; ({\rm fixed})\\
                                                              \end{array}\right\}$   
                           \\
$(R/10^3 \;{\rm km})^2$\tablenotemark{\dagger}       &  3 -- 2100000     &9.2$\pm^{2.9}_{2.2}$  \\
$L_{bol}$ [$10^{36}$ erg/s]\tablenotemark{\dagger}  &  15 --   630000     & $38\pm^{12}_{10}$ \\
$\chi^2$/dof                         & 21.9 (31)              &23.9 (32)\\\hline
\end{tabular}
\tablenotetext{\dagger}{The distance is assumed to be 55 kpc.}
\tablecomments{Absorption edge of the form $\exp(-\tau (E/E_{\rm edge})^{-3})$ (for $E>E_{\rm edge}$)
 is assumed.}
\end{center}
\end{table}
%

\begin{table}
\caption{CAL87 LTE model fit: log $g$=9, LMC abundance.}\label{cal87_lte_fit_table}
\begin{tabular}{lccccc} \\\hline\hline
                                     & All                   &\multicolumn{2}{c}{Eclipse} & \multicolumn{2}{c}{Out of Eclipse}  \\\hline
$N_H [10^{21}$cm$^{-2 }]$            & 2.0$\pm0.7$           &  fixed     &  2.5$\pm$0.2  &     fixed       &   1.95$\pm$0.09     \\
$T_{eff}$ [eV]                       & 75$\pm$1              & \multicolumn{2}{c}{fixed}  &    \multicolumn{2}{c}{fixed} \\
$(R/10^3 \;{\rm km})^2$ \tablenotemark{\dagger}    & 1.6$\pm^{0.9}_{0.6}$  & 1.34$\pm$0.11&   fixed     &     1.67$\pm$0.08  & fixed      \\
$L_{bol}$ [$10^{36}$ erg/s] \tablenotemark{\dagger}& 6.6$\pm^{3.6}_{2.5}$  & 5.5$\pm$0.5 &    fixed     &      6.8$\pm$0.3   &  fixed       \\
$\chi^2$/dof                         &39.0/32                &  34.7/34  & 33.3/34        &    29.4/34         & 29.0/34    \\\hline
\end{tabular}
\tablenotetext{\dagger}{The distance is assumed to be 55 kpc.}
\end{table}

\begin{table}
\caption{CAL87 NLTE model fit: log $g$=9, LMC abundance.}\protect\label{cal87_nlte_fit_table}
\begin{tabular}{lccccc} \\\hline\hline
                                      & All                   &\multicolumn{2}{c}{Eclipse}& \multicolumn{2}{c}{Out of Eclipse}  \\\hline
$N_H [10^{21}$cm$^{-2 }]$             & 2.0$\pm0.6$           &  fixed       &2.4$\pm^{0.2}_{0.1}$&   fixed                &   $1.95\pm0.09$      \\
$T_{eff}$ [eV]                        & 58$\pm$1              & \multicolumn{2}{c}{fixed} & \multicolumn{2}{c}{fixed} \\
$(R/10^3 \;{\rm km})^2$ \tablenotemark{\dagger}     & 4.5$\pm^{2.4}_{2.7}$  & 3.7$\pm$0.3  &      fixed & 4.7$\pm$0.2       & fixed      \\
$L_{bol}$ [$10^{36}$ erg/s] \tablenotemark{\dagger} &6.6$\pm^{3.4}_{4.0}$   & 5.4$\pm$0.5  &  fixed     & 6.9$\pm$0.3       & fixed        \\
$\chi^2$/dof                          & 23.1/32               &  34.6/34     & 33.9/34    &    14.6/34             & 14.3/34    \\\hline
\end{tabular}
\tablenotetext{\dagger}{The distance is assumed to be 55 kpc.}
\end{table}

\begin{table}
\caption{CAL87 average spectrum model fit with the cosmic abundance and different surface gravities.}
\protect\label{cal87_abundance}
\begin{tabular}{lcccc} \\\hline\hline
                                     & \multicolumn{2}{c}{LTE model}                   & \multicolumn{2}{c}{NLTE model}                   \\
log $g$ (CGS)                         & 9                    & 10                      & 9                       & 10  \\\hline
$N_H [10^{21}$cm$^{-2 }]$            & 2.2$\pm0.9$           & 1.8$\pm{0.8}$           & 2.2$\pm$0.7             & 1.2$\pm^{1.0}_{0.5}$     \\
$T_{eff}$ [eV]                       & 75$\pm^{2}_{1}$       & 89$\pm^2_1$             & 65$\pm1$                & 79$\pm$2\\
$(R/10^3 \;{\rm km})^2$ \tablenotemark{\dagger}    & 2.1$\pm1.2$           & 0.71$\pm^{0.50}_{0.33}$ & 3.3$\pm^{2.7}_{1.8}$    & 0.66$\pm^{0.71}_{0.20}$     \\
$L_{bol}$ [$10^{36}$ erg/s] \tablenotemark{\dagger}& $8.6\pm4.4$         & 5.8$\pm^{4.0}_{2.4}$ & 7.7$\pm^{5.5}_{4.0}$ & 3.4$\pm^{2.9}_{0.5}$       \\
$\chi^2$/dof                         & 41.4/32               & 42.8/32                 & 25.2/32                 & 28.7/32                 \\\hline
\end{tabular}
\tablenotetext{\dagger}{The distance is assumed to be 55 kpc.}
\end{table}

\begin{table}
\begin{center}
\caption{RXJ0925.7--4758 blackbody (plus edges)  model fit}\label{rxj_bb_edge_table}\label{rxj_bb_table}
\begin{tabular}{lcccc}\hline\hline
$N_H [10^{22}$cm$^{-2 }]$            & 2.0 & 1.14  & 1.06$\pm0.13$           &$1.09\pm0.11$\\
$T$ [eV]                             & 45  &  88   & 92$\pm^{8}_{10}$        &$ 88 \pm^{10}_{6}$\\
Edge1$\left\{ \begin{array}{l}
        E_{edge} \; [{\rm keV}]\\ 
        \tau
      \end{array}\right.$            & --- &    $\left.\begin{array}{l}
                                            0.86 \\ 
                                            1.17 \\
                                             \end{array}\right.$ &   $\left.\begin{array}{l}
                                                                   0.90\pm^{0.02}_{0.03}   \\ 
                                                                   0.81\pm0.3\\
                                                                  \end{array}\right.$   &  ~~~~~~---    $\left.\begin{array}{c}
                                                                                                             \\
                                                                                                              \\
                                                                                                          \end{array}\right\}$                  \\
Edge2$\left\{ \begin{array}{l}
        E_{edge} \; [{\rm keV}]\\ 
        \tau
      \end{array}\right.$            & --- &$\left.\begin{array}{l}
                                       1.01  \\ 
                                       2.35 \\
                                        \end{array}\right.$     &  $\left.\begin{array}{l}
                                                                  1.02\pm0.02   \\ 
                                                                  1.64\pm^{0.3}_{0.4}  \\
                                                             \end{array}\right.$     &  $\left.\begin{array}{l}
                                                                                       1.01\pm^{0.01}_{0.02}\\ 
                                                                                       1.74\pm0.3  \\
                                                                                   \end{array}\right\}$   
                           \\
Edge3$\left\{ \begin{array}{l}
        E_{edge} \; [{\rm keV}]\\ 
        \tau
      \end{array}\right.$            & --- &$\left.\begin{array}{l}
                                       1.31  \\ 
                                       1.18 \\
                                        \end{array}\right.$     & $\left.\begin{array}{l}
                                                               1.36\pm^{0.07}_{0.05}   \\ 
                                                               0.83\pm^{0.7}_{0.6} \\
                                                                        \end{array}\right.$     & $\left.\begin{array}{l}
                                                                                           1.36\pm^{0.09}_{0.07}   \\ 
                                                                                           0.68\pm0.6 \\
                                                                                             \end{array}\right\}$   
                           \\
Oxygen abundance\tablenotemark{a}    &    1 (fixed)             &  1(fixed)          &0.61$\pm$0.1                  & $0.65\pm$0.1\\ 
Neon   abundance\tablenotemark{a}    &    1 (fixed)             &  1(fixed)          &1 (fixed)                     & $2.3\pm0.5$\\                            
$R^2$ [km$^2$] \tablenotemark{\dagger}              & $3.34\times10^{10}$      &$22.9 \times 10^4$  &5.2$\pm^{19}_{4}\times10^4$   & $9.60\pm^{26}_{7}\times10^4$\\
$L_{bol}$ [erg/s] \tablenotemark{\dagger}           &  $1.8\times10^{40}$      & $1.8\times10^{36}$ &4.8$\pm^{9}_{3}\times10^{35}$ & $7.5\pm^{13}_{4}\times10^{35}$\\
$\chi^2$/dof                         & 605 (75)  &102.8 (69)    &49.2 (68)                                          &50.8 (69)\\\hline
\end{tabular}
\tablenotetext{\dagger}{The distance is assumed to be 1 kpc.}
\tablenotetext{a}{Abundances for the interstellar absorption,
relative to the cosmic abundances [O]/[H]=7.39$\times10^{-4}$  and 
[Ne]/[H]=$1.38\times10^{-4}$ by Anders and Ebihara (1982). 
The same definition for the following Tables.
}
\tablecomments{Absorption edge of the form $\exp(-\tau (E/E_{\rm edge})^{-3})$ (for $E>E_{\rm edge}$)
 is assumed.}
\end{center}
\end{table}
%

\clearpage

\begin{table}
\begin{center}
\caption{RXJ0925.7--4758 LTE and NLTE model fit results.}\label{rxj_lte_nlte_table}
\begin{tabular}{lcccccc}	\hline\hline
                       & \multicolumn{3}{c}{LTE} & \multicolumn{3}{c}{NLTE}\\\hline
log $g$ (CGS)          & 9   &   10  & 10  & 9   & 10 & 10\\
$N_H$ [$10^{21}$ cm$^{-2}$]
                       &21.1  & 16.5 & 7.1$\pm$0.6 & 8.7  &8.5&6.9$\pm^{0.6}_{0.2}$\\
Oxygen abundance  &0.92 & 0.90 &0.56$\pm$0.14&0.65  &0.62  & 0.61$\pm$0.15      \\   
Neon abundance    &0.33 &  0   &0.56$\pm^{0.9}_{0.56}$    &2.2   &0.26  & 4.0$\pm^{0.3}_{0.5}$    \\   
Edge 
      $\left\{ \begin{array}{l}
        E_{edge} \; [{\rm keV}]\\ 
                       \tau
           \end{array}\right.$   & ---    &   ---     &   $
                                   \begin{array}{c}
                                    1.02\pm0.01\\
                                    2.1\pm0.3 \\
                                   \end{array}$
                                                       & --- & --- & 
                                                                     $\left.\begin{array}{c}
                                                                      1.01\pm0.01\\
                                                                      1.72\pm^{0.10}_{0.16}\\
                                                                       \end{array}\right\}$\\      
$T_{eff}$ [eV]         & 70  & 86                       & 117$\pm^4_2$    &  74    &92 &  103\tablenotemark{a}  \\
$R^2$ [km$^2$] \tablenotemark{\dagger}&$2.6\times10^8$ &$3.3\times10^6$&$1.26\pm^{0.28}_{0.35}\times10^3$  &  $2.7\times10^4$   &$7.7\times10^3$ &$2.6\times10^3$      \\                                   
$L_{bol}$ [ergs s$^{-1}$] \tablenotemark{\dagger}     
                       &$8\times10^{38}$&$2.3\times10^{37}$ &$3.1\times10^{34}$&$1\times10^{35}$&$7\times10^{34}$&$3.8\times10^{34}$\\                                   
$\chi^2$ (dof)         & 288 (73)  & 288 (73)   & 79.2 (71)    & 393 (73)   & 343 (73)    & 51.7 (73)\\\hline                                   
\end{tabular}
\tablenotetext{\dagger}{The distance is assumed to be 1 kpc.}
\tablenotetext{a}{103 eV (= $1.2 \times 10^6$ K) is the highest available temperature in the model.}
\end{center}
\end{table}

\begin{table}
\begin{center}
\caption{RXJ0925.7--4758 LTE plus Raymond and Smith (1977) plasma model fit result}\label{rxj_lte_ray_table}
\begin{tabular}{lc}	\hline
log $g$ (CGS)          & 10  \\
$N_H$ [$10^{21}$ cm$^{-2}$]
                       &12.7\\
Oxygen abundance  &0.67     \\   
Neon abundance    &0.68     \\   
$T_{eff}$ [eV] &           94              \\
$R^2$ [km$^2$] \tablenotemark{\dagger}&$7 \times 10^4$     \\                                   
$T$ [eV]\tablenotemark{a}            & 103            \\
$N$\tablenotemark{b}             &      $1.5 \times 10^{53}$                   \\
$\chi^2$ (dof)         & 144 (71)\\\hline                                   
\end{tabular}
\tablenotetext{\dagger}{The distance is assumed to be 1 kpc.}
\tablenotetext{a}{Plasma temperature.}
\tablenotetext{b}{The plasma emission measure $n_e n_H V$ [cm$^{-3}$]   at 1 kpc, 
where $n_e$, $n_H$ and $ V$ are electron density, hydrogen density and volume, respectively.}
\end{center}
\end{table}

\clearpage

%

%

%

%

\clearpage

%

\clearpage

\clearpage

\clearpage

\clearpage

\begin{figure}
\centerline{
\psfig{figure=f11b.eps,%
      height=9.cm,bbllx=0.cm,bblly=4.cm,bburx=21.cm,bbury=21cm,%
      angle=0,clip=}}
\centerline{
\psfig{figure=f11a.eps,%
      height=9.cm,bbllx=0.cm,bblly=4.cm,bburx=21.cm,bbury=21cm,%
      angle=0,clip=}
}
\caption{NLTE model fit for RXJ0925.7--4758 with the surface gravity
 $10^{10}$ cm s$^{-2}$ with an absorption edge at 1.01 keV.
The oxygen and neon abundances in the interstellar absorption are made
free parameters  (see the text).
Left: Observed spectrum and best-fit model convolved with the detector response.
The fitting residual is also shown.
Right: The best-fit model before and after  the interstellar absorption.
}\label{rxj_nlte_log_10_edge}
\end{figure}

\begin{figure}
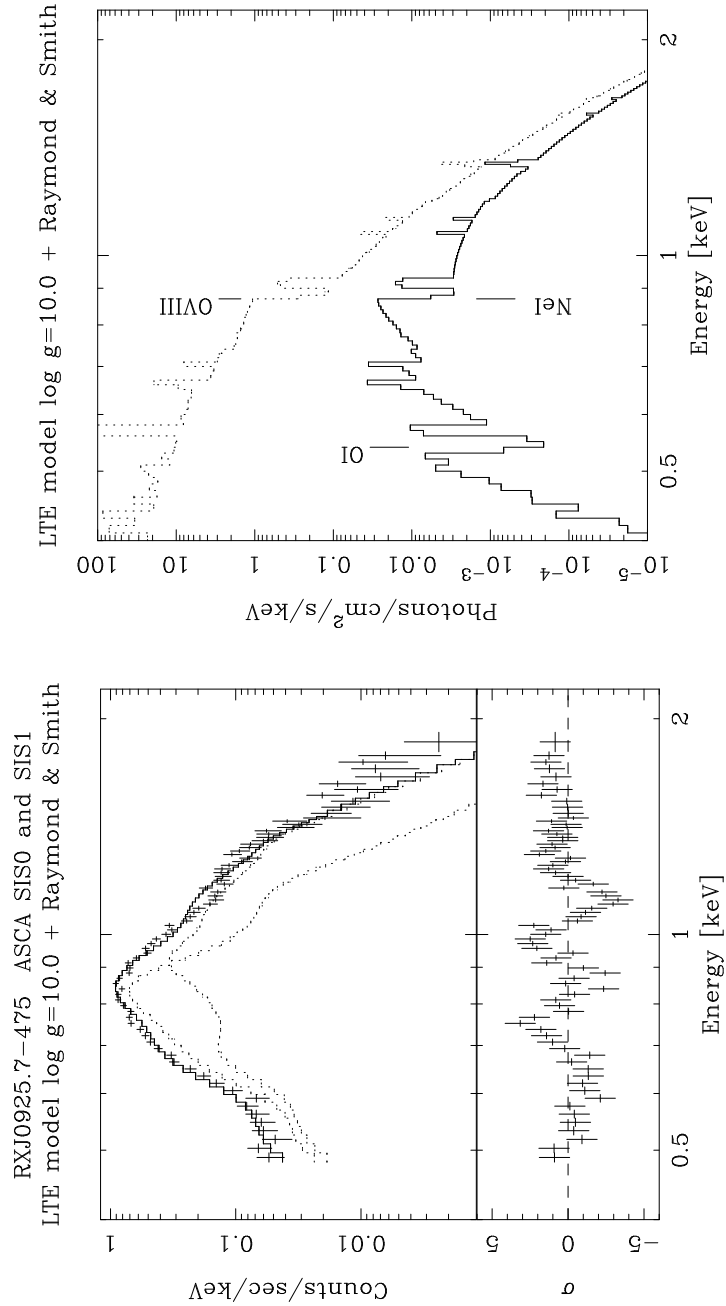

\centerline{
\psfig{figure=f12b.eps,%
      height=9.cm,bbllx=0.cm,bblly=4.cm,bburx=21.cm,bbury=21cm,%
      angle=0,clip=}}
\centerline{
\psfig{figure=f12a.eps,%
      height=9.cm,bbllx=0.cm,bblly=4.cm,bburx=21.cm,bbury=21cm,%
      angle=0,clip=}
}
\caption{LTE model (surface gravity  $10^{10}$ cm s$^{-2}$)
plus Raymond \& Smith plasma model  
fit for RXJ0925.7--4758.
The oxygen and neon abundances in the interstellar absorption are made
free parameters  (see the text).
Left: Observed spectrum and best-fit model convolved with the detector response.
The fitting residual is also shown.
Right: The best-fit model before and after  the interstellar absorption.
}\label{rxj_lte_ray}
\end{figure}

%
\begin{figure}
\centerline{
\psfig{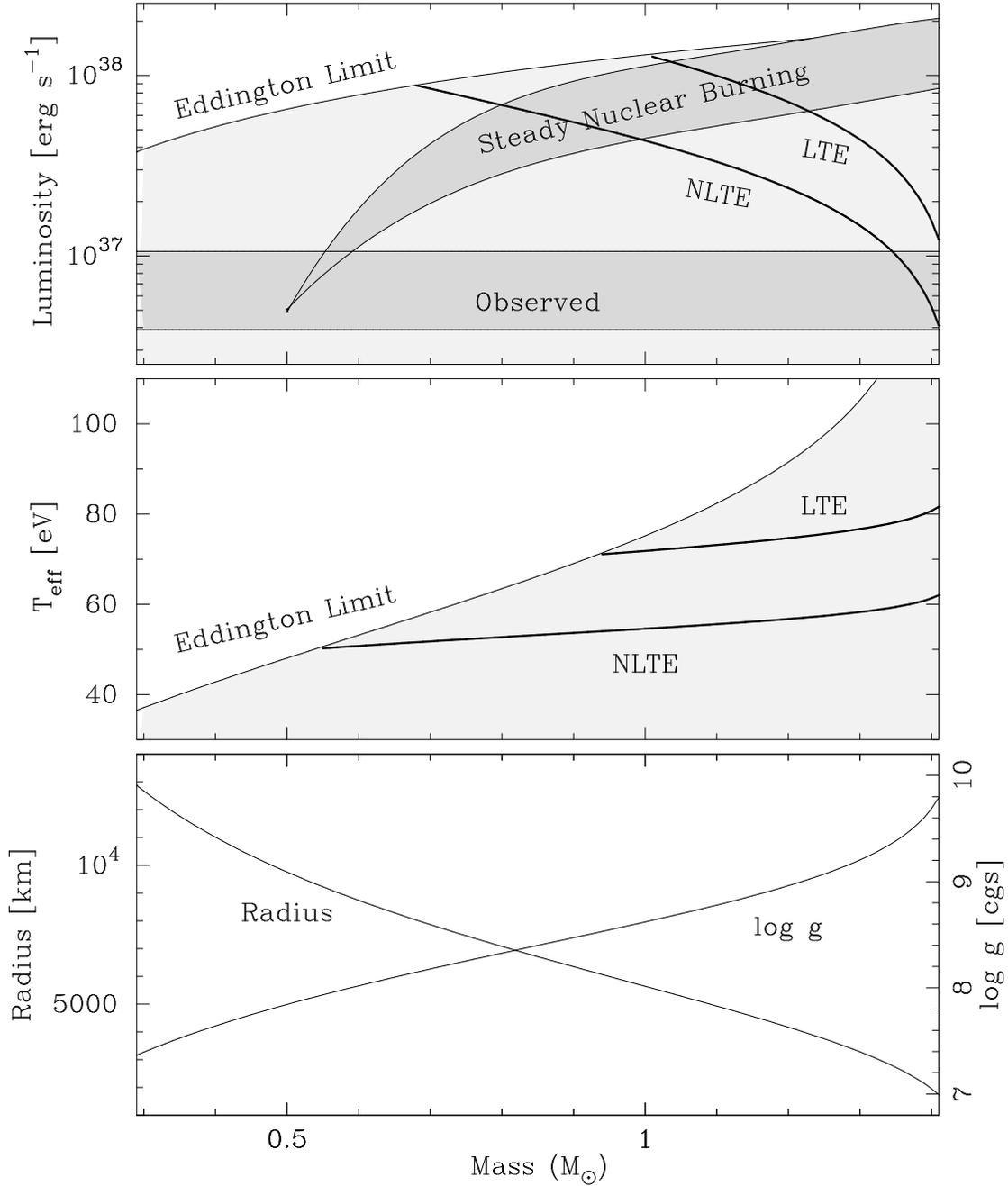}
}
\caption{
Radius, surface gravity and the maximum effective temperatures at the Eddington luminosity 
of a white-dwarf are  calculated as  functions  of the mass.
In the bottom panel, 
the mass-radius relationship of Pringle and Webbink (1975) is assumed,
based on which the surface gravity and the temperatures at the Eddington 
luminosities  are calculated.
In the middle and top panels, the best-fit temperatures and luminosities
of CAL87 determined  using the LTE and NLTE models are indicated by thick lines. 
The steady nuclear burning zone of the SSS model by 
van den Heuvel et al.\ (1996) is shown in the top figure,
as well as the range of the observed CAL87 luminosity
assuming isotropic emission.
See texts for further explanations.
}\label{white_dwarf_figure}
\end{figure}
%
%
\begin{figure}
\centerline{
\psfig{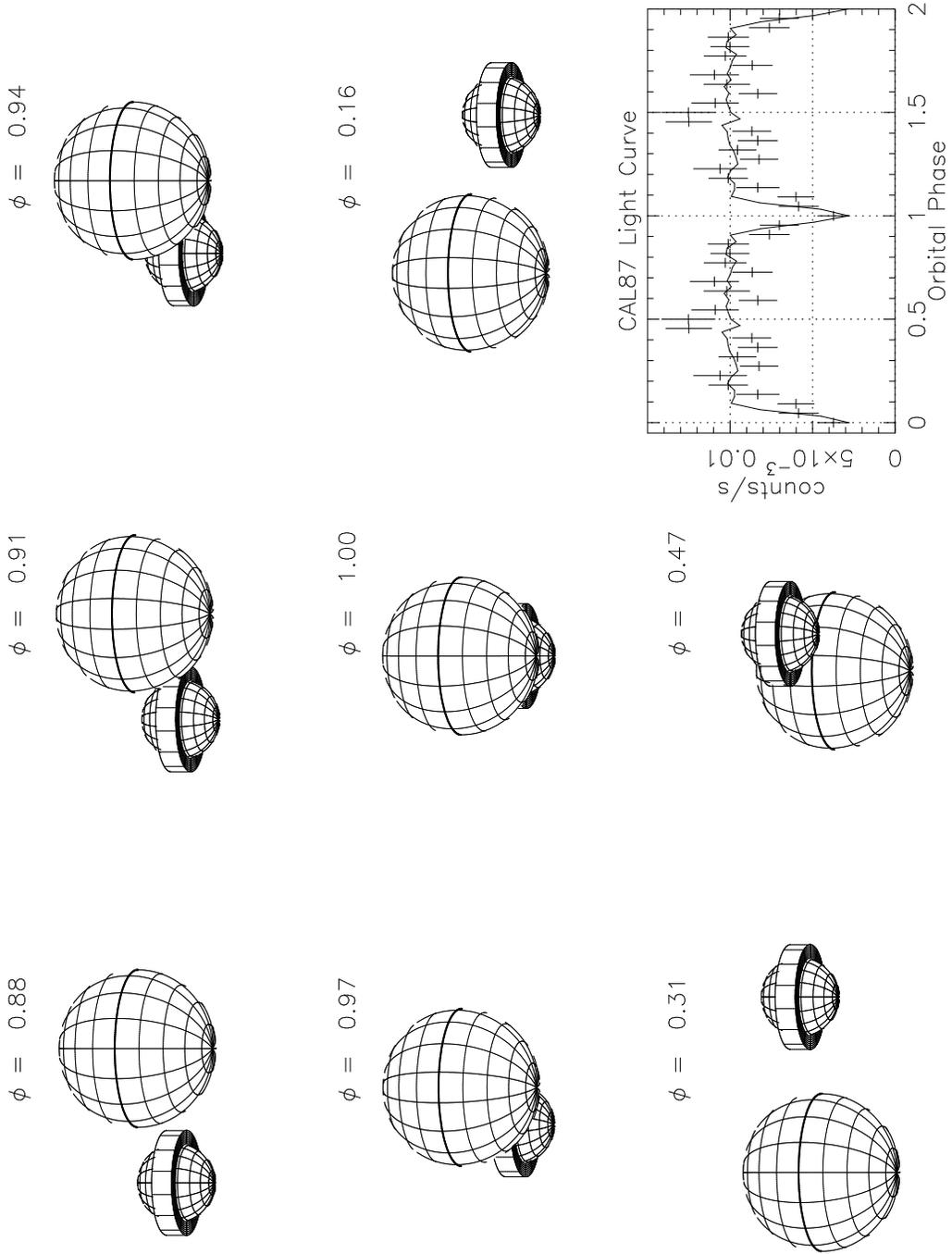}
}
\caption{
Simulation of the CAL87 X-ray orbital light curve with the Accretion
Disk Corona (ADC) model.
The smaller sphere represents the X-ray emitting ADC, and the
surrounding ``band'' is  the accretion disk to block the X-rays from
the white dwarf and ADC. X-ray flux is assumed to be proportional to
the projected ADC surface area which is not blocked by either companion star 
(filling its Roche lobe) or accretion disk.
The accretion disk radius is assumed to be  0.8 times the white dwarf 
Roche lobe radius, and the ADC radius is  0.75 times the disk radius. Inclination angle is $73^\circ$.  For other orbital parameters used, see texts.
}\label{light_curve1}
\end{figure}

\end{document}